\documentclass[aip,jcp
amsmath,amssymb,
reprint,%
]{revtex4-2}

\usepackage{dcolumn}% Align table columns on decimal point
\usepackage[T1]{fontenc}
\usepackage{mathptmx}
\usepackage{etoolbox}
\usepackage{bm}
\usepackage{url}
\usepackage{graphicx}
\usepackage{amsmath}
\usepackage{amssymb}
\usepackage[utf8]{inputenc}
\usepackage{ulem}
\usepackage{enumerate}
\usepackage{float}
\usepackage[usenames]{color}
\usepackage[sort&compress]{natbib}
\usepackage{xr}
\DeclareUnicodeCharacter{2212}{\textendash} % defining dash character for bib entries

\draft % marks overfull lines with a black rule on the right

\begin{document}

% Use the \preprint command to place your local institutional report number 
% on the title page in preprint mode.
% Multiple \preprint commands are allowed.
%\preprint{}

\title{Scaling Law of Quantum Confinement in Single-Walled Carbon Nanotubes} %Title of paper

% repeat the \author .. \affiliation  etc. as needed
% \email, \thanks, \homepage, \altaffiliation all apply to the current author.
% Explanatory text should go in the []'s, 
% actual e-mail address or url should go in the {}'s for \email and \homepage.
% Please use the appropriate macro for the type of information

% \affiliation command applies to all authors since the last \affiliation command. 
% The \affiliation command should follow the other information.
%\author{}
%\email[]{Your e-mail address}
%\homepage[]{Your web page}
%\thanks{}
%\altaffiliation{}
%\affiliation{}

\author{Benjamin Eller}
\affiliation{Institute for Physical Sciences and Technology, University of Maryland, College Park, MD 20742, USA}
\author{Charles W. Clark}
\affiliation{Joint Quantum Institute, National Institute of Standards and Technology and the University of Maryland, Gaithersburg, MD 20899, USA}
\author{YuHuang Wang}
\affiliation{Department of Chemistry and Biochemistry, University of Maryland, College Park, MD 20742, USA}

% Collaboration name, if desired (requires use of superscriptaddress option in \documentclass). 
% \noaffiliation is required (may also be used with the \author command).
%\collaboration{}
%\noaffiliation

\date{\today}

\begin{abstract}
Quantum confinement significantly influences the excited states of sub-10 nm single-walled carbon nanotubes (SWCNTs), crucial for advancements in transistor technology and the development of novel opto-electronic materials such as fluorescent ultrashort nanotubes (FUNs). However, the length dependence of this effect in ultrashort SWCNTs is not yet fully understood in the context of the SWCNT exciton states. Here, we conduct excited state calculations using time-dependent density functional theory (TD-DFT) on geometry-optimized models of ultrashort SWCNTs and FUNs, which consist of ultrashort SWCNTs with $sp^3$ defects. Our results reveal a length-dependent scaling law of the $E_{11}$ exciton energy that can be understood through a geometric, dimensional argument, and which departs from the length scaling of a 1D particle-in-a-box. We find that this scaling law applies to ultrashort (6,5) and (6,6) SWCNTs, as well as models of (6,5) FUNs. In contrast, the defect-induced $E_{sp^3}$ transition, which is redshifted from the $E_{11}$ optical gap transition, shows little dependence on the nanotube length, even in the shortest possible SWCNTs. We attribute this relative lack of length dependence to orbital localization around the quantum defect that is installed near the SWCNT edge. Our results illustrate the complex interplay of defects and quantum confinement effects in ultrashort SWCNTs and provide a foundation for further explorations of these nanoscale phenomena. 
\end{abstract}
% insert suggested PACS numbers in braces on next line

\maketitle %\maketitle must follow title, authors, abstract and \pacs

% Body of paper goes here. Use proper sectioning commands. 
% References should be done using the \cite, \ref, and \label commands
\section{Introduction}
Ultrashort SWCNTs, particularly those shorter than the exciton diffusion length (< 90 nm \cite{Cognet2007}) and approaching the sub-10 nm regime, are central components for devices such as carbon nanotube transistors  \cite{Cao2005,Franklin2010,Franklin2012,Cao2015,Qiu2017,Cao2017,Janissek2021,Franklin2022}, as well as light emission sources for near-IR sensing, imaging, and single-photon based quantum information applications enabled by $sp^3$ defect-induced fluorescence\cite{Piao2013,He2017,Danne2018,Li2019,Brozena2019,Borel2023}. Understanding the length dependence of excited state energies in ultrashort SWCNTs is necessary for interpreting experiments involving electrical transport, photon emission and absorption measurements. In the case of optical experiments, the energies of the exciton states are most relevant as these states are chiefly responsible for the optical properties of SWCNTs\cite{Wang2005}. Significant finite-size effects, such as those observed in quantum dots due to quantum confinement, also occur in SWCNTs. Numerous studies have investigated finite-size (i.e. length) effects in SWCNTs, both experimentally\cite{Venema1999,Heller2004,Fagan2007,Sun2008,Morimoto2014,Gao2015} and theoretically\cite{Zhu1998,Rochefort1999,Saito1999,Wu2000,Compernolle2003,Zhou2004,Lu2004,Chen2004,Yumura2004a,Yumura2004b,Bettinger2004,Yumura2006,Okada2007,Rajan2008,Silva2016,Eller2022}. Despite this, our understanding of such length effects on the optical properties of SWCNTs has been limited to relatively long nanotubes. The quantum confinement effect in sub-10 nm SWCNTs is crucial to explore because the greatest blueshifts should occur in this range due to their inverse length dependence, such as can be seen\cite{Gao2015} employing the typical 1D particle-in-a-box picture in which $\Delta E_{11}\sim L^{-2}$. However, an equation that captures the length dependent energy shift of the exciton transitions in ultrashort SWCNTs, similar to the Brus equation\cite{Brus1984} for spherical quantum dots, remains elusive. 
\begin{figure*} 
	\centering
	\begin{minipage}{0.5\textwidth}
		\raggedright
		\includegraphics[scale=0.1]{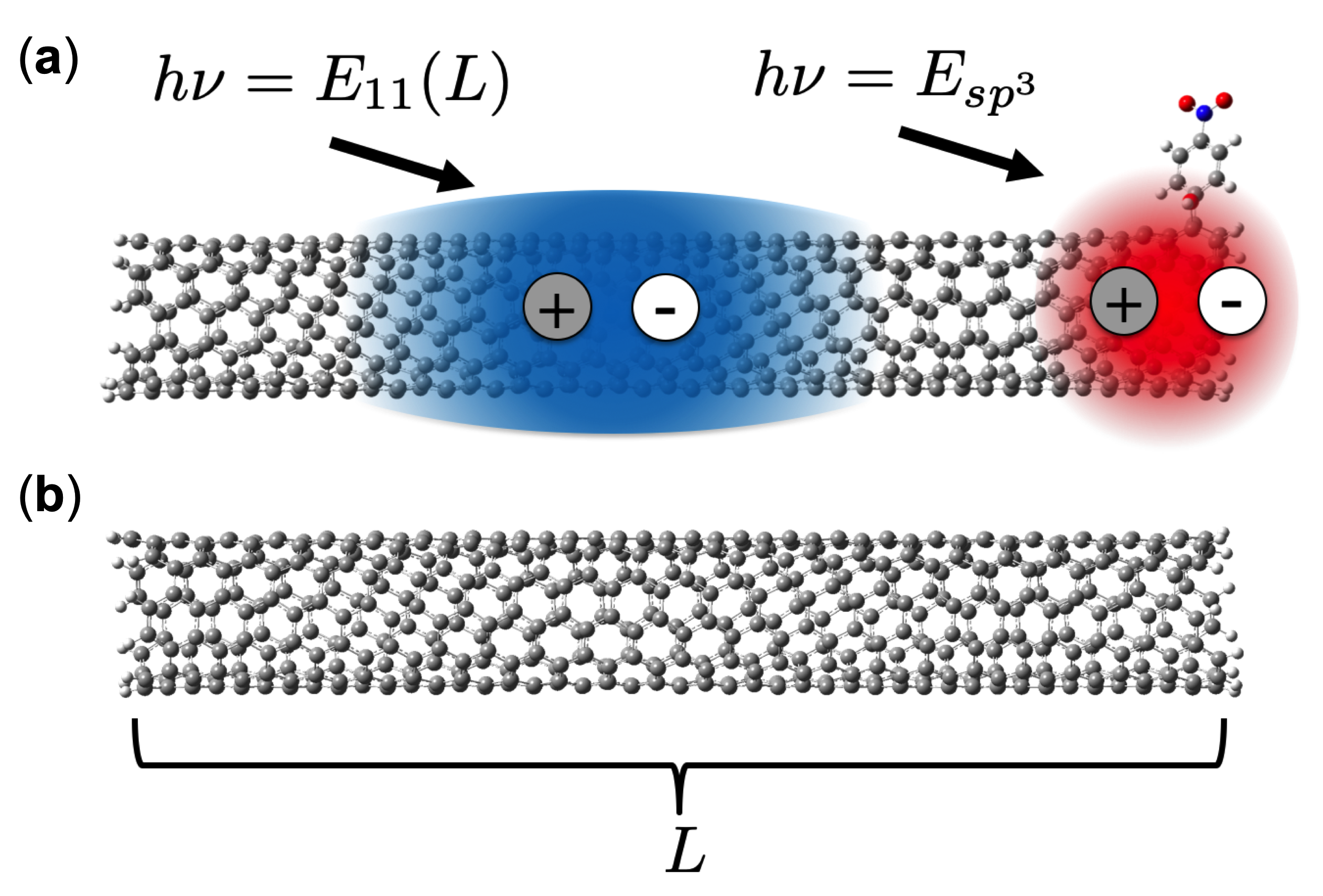}
		\label{fig1a}
	\end{minipage}%
	\begin{minipage}{0.5\textwidth}
		\raggedleft
		\includegraphics[scale=0.1]{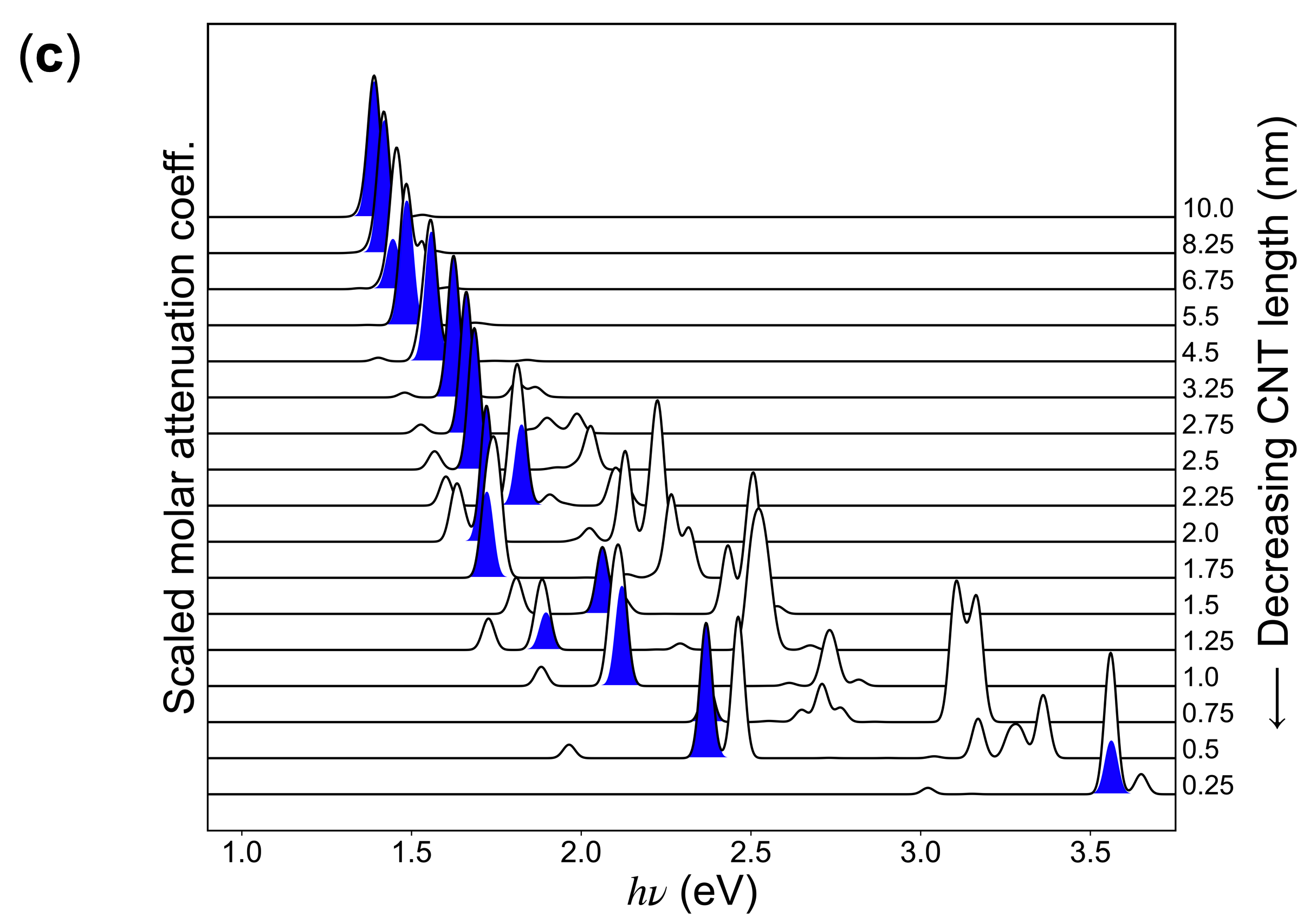}
		\label{fig1b}
	\end{minipage}
	\caption{\textbf{Quantum confinement in ultrashort SWCNTs.} \textbf{(a)} Molecular model of an edge-functionalized (6,5) SWCNT, a simplified example of a FUN. Photons of energy $h\nu$, with $\nu$ being the frequency, are absorbed by the SWCNT systems creating excited electron-hole pairs. \textbf{(b)} Molecular model of a pristine (6,5). Both models have the same length $L$ of approximately 5.5 nm, and are passivated with hydrogen atoms at the edges. \textbf{(c)} Calculated absorption spectra of (6,5) SWCNTs decreasing in length (from top to bottom), showing quantum confinement effects. The spectra are scaled by normalizing each one to the maximum value of the molar attenuation coefficient over the entire dataset, and offset in the vertical axis for clarity. The $E_{11}$ transitions selected via an NTO analysis (Fig.\,\ref{supp-NTO_wave}) to extract the scaling are highlighted in blue (color online). As the SWCNT length decreases, the transitions increase in energy.}
	\label{fig1}
\end{figure*}   

The difficulty arises from the challenge of obtaining and manipulating individualized ultrashort SWCNT samples with precisely controlled lengths to measure resulting blueshifts in the excitonic transitions. Further complications arise as photoluminescence (PL) is typically quenched in ultrashort SWCNTs \cite{Cherukuri2012,Danne2018}, which restricts their use as near-IR light sources. However, two recent advances in SWCNT chemistry have made it possible to probe the PL of SWCNTs in the ultrashort regime, renewing interest in ultrashort SWCNTs as near-IR emitters. The first advance is the application of SWCNT $sp^3$ defect chemistry\cite{Piao2013}, known to significantly enhance the PL of SWCNTs, to ultrashort SWCNTs to create fluorescent ultrashort nanotubes (FUNs). In FUNs, $sp^3$ defects placed near the edges of ultrashort SWCNTs activate the normally quenched PL of $E_{11}$ excitons, in addition to creating bright defect exciton PL\cite{Danne2018}. A simplified model of a FUN is shown in Fig.\,\ref{fig1}a, which consists of a SWCNT with an $sp^3$ defect pair near one edge. The second advance is an improved technique for creating FUNs called defect-induced chemical etching (DICE)\cite{Li2019}, which is enabled by $sp^3$-functionalization of SWCNTs. This technique uses hydrogen peroxide to produce clean cuts at the $sp^3$ defect sites, leaving the luminescent $sp^3$ defects behind at the edges of the thus shortened nanotubes. This method provides a mechanism for molecular tunability of the length distributions of FUNs based on the linear $sp^3$ defect density $\rho_{sp^3}$ of the initially functionalized SWCNTs, since the expected average lengths of the cut FUNs should be $\bar{L}\sim\rho^{-1}_{sp^3}$. In principle, the DICE method allows for precise access to fluorescent ultrashort SWCNTs. 

Inspired by these developments, we revisit the question of the length dependence of the $E_{11}$ exciton energy in the context of ultrashort SWCNTs and FUNs. Existing theoretical studies typically investigate the confinement effect by considering the impact of finite lengths on the ground-state electronic structure, usually through either a tight-binding or DFT-based approach \cite{Zhu1998,Rochefort1999,Wu2000,Lu2004,Chen2004,Yumura2006,Okada2007}. While this provides valuable insight into the electronic structure of ultrashort SWCNTs, what is necessary for studying the confinement energy scaling in these systems is an excited state approach that incorporates electron-hole interactions, as the optical resonances in SWCNTs are excitonic in nature\cite{Wang2005}. 

Time-dependent density functional theory (TD-DFT) offers a promising method for calculating the excited electronic states of molecular SWCNT models across a wide range of lengths, from the smallest definable segments to models that approximate the infinite-length limit. TD-DFT methods using the B3LYP approximation to the exchange-correlation energy have successfully provided insights into low-lying exciton states in SWCNT models with lengths $\approx 10$ nm\cite{Kilina2009,Kilina2012}. At these lengths the calculated absorption spectra qualitatively reflect the infinite-length limit, and in pristine (6,5) SWCNTs the $7^{\rm th}$ excited state can be identified as the bright $E_{11}$ exciton when using the B3LYP approximation. However, previous TD-DFT studies\cite{Zhou2004,Silva2016,Eller2022} examining confinement effects on absorption spectra in ultrashort SWCNTs did not identify a clear length dependence of the lowest bright exciton energy. 

Here, we present TD-DFT calculations of the $E_{11}$ exciton state in pristine (6,5) SWCNTs. The calculations are performed on molecular models of (6,5) SWCNTs spanning lengths from roughly 0.25−10 nm. An example model is shown in Fig.\,\ref{fig1}b, depicting a finite-length (6,5) SWCNT of length $L\approx5.5$ nm. 

Our results show a scaling law for the $E_{11}$ energy length dependence in ultrashort SWCNTs: 
\begin{equation}
\Delta E_{11} = E_{11}(L)-E_{\infty} \sim L^{-1/2}. 
\end{equation}
This law differs from the conventional 1D particle-in-a-box picture sometimes applied to understand the effect of quantum confinement on excitons in SWCNTs\cite{Gao2015}, and contrasts with the confinement effect in spherical quantum dots\cite{Brus1984}. To explain this unexpected length dependence, we make a dimensional argument that combines the strong Coulomb interaction between an electron and hole in SWCNTs and the nanotube geometry hosting the excited electron-hole pair. 

We show that our scaling law also applies to the bright, low-energy excited states that have been previously seen in ultrashort (6,6) SWCNTs \cite{Eller2022}. We also connect the scaling law to models of (6,5) FUNs with edge-$sp^3$ defects, where we see both the $E_{11}$ as well as a redshifted defect-localized transition that we label $E_{sp^3}$. The observation of an $sp^3$ defect-induced transition redshifted from the $E_{11}$ mirrors what is observed in experimental probes of the excited states of FUNs via PL \cite{Danne2018,Li2019}. 

The presence of our scaling law in both semiconducting and metallic chiralities, as well as in SWCNTs with $sp^3$ defects, suggests a broad applicability of the dimensional argument to ultrashort SWCNT systems. As demonstrated by Li et. al.\cite{Li2019}, the DICE chemistry provides the ability to synthesize FUNs with chemically controlled lengths, which constitute a promising class of SWCNT quantum dot molecules. These materials will be useful for a variety of applications ranging from bioimaging to quantum information science, as well as for the measurement of length-dependent properties of ultrashort SWCNTs, e.g. the length-dependence of the $E_{11}$ energy seen in our calculations here.

\section{Methods}
All TD-DFT calculations in this paper were performed with Gaussian 16 computational chemical software\cite{g16}. The molecular models used in this study were generated in a manner inspired by pioneering applications of the Clar theory of polycyclic hydrocarbons to SWCNTs\cite{Matsuo2003,Ormsby2004,Baldoni2007,Balaban2009}. The Clar model approach is advantageous for studying molecular models of SWCNTs compared to using unit cells\cite{Baldoni2007}, particularly for chiral nanotubes such as (6,5) SWCNTs, where the Clar cell is roughly a quarter of a nanometer long compared to the unit cell that is about 4 nm long. The length of any of our finite SWCNT models can be expressed as $x=L/L_0$ (in cu) or $L$ (in nm), where $x$ is essentially dimensionless as it is an integer used to control the number of screw-axis operations used to generate the models, and $L_0$ is a chirality-specific natural length scale of the models. For the (6,5) and (6,6) chiralities studied in this work, $L^{(6,5)}_0\approx 0.25$ nm and $L^{(6,6)}_0\approx 0.125$ nm.  

For the molecular models of (6,5) FUNs, the defect installed near the edge of the SWCNTs can be denoted as (4-nitrobenzene, OH, ortho-90)\cite{Eller2022}. All results presented on FUNs in this paper are referring to these models with a single $sp^3$ defect pair near one edge of an ultrashort (6,5) SWCNT, which serve as idealized examples of FUNs. Additional details of the methods used in this study are provided in the Supplemental Materials, including details of our computations, coordinates for example models, a schematic depiction of the screw axis operation used to generate the SWCNT models, as well as the $sp^3$ defect orientation in the FUN models (Fig.\,\ref{supp-model_schematic}).

\section{Results and Discussion}
We used the first 15 excited states of the geometry-optimized models calculated by TD-DFT to construct the electronic absorption spectra with the excited state energies and oscillator strengths. The absorption spectra for all pristine (6,5) SWCNTs studied here are displayed in Fig.\,\ref{fig1}c, where we see the effect of quantum confinement blueshifting the excited state energies as the length is decreased. We analyzed these spectra to identify the $E_{11}$ transitions for different lengths. In pristine (6,5) SWCNTs with $L \sim 10$ nm, the $7^{\rm th}$ excited state can be identified as the bright $E_{11}$ transition\cite{Kilina2009}. For models longer than approximately 2.5 nm ($x=10$ cu), the $E_{11}$ transition is straightforward to identify by inspecting the absorption spectra, as there is only one significant peak, the $E_{11}$ state. In models shorter than 2.5 nm, this peak splits into multiple transitions, making the assignment less clear. 

In order to assign the $E_{11}$ in SWCNTs shorter than 2.5 nm, we perform a natural transition orbital (NTO) analysis\cite{Martin2003} of the $E_{11}$ excited state in the 10 nm (6,5) SWCNT. Natural transition orbitals provide an orbital description of excited states that is more intuitive than using Kohn-Sham molecular orbitals. We find that in the 10 nm (6,5) SWCNT, the $E_{11}$ exhibits the waveform shown in Fig.\,\ref{supp-NTO_wave}. Typically, two relevant NTO pairs (electron-hole) are involved in the $E_{11}$ transitions, usually with similar contributions to the excited state. The most important NTO pair consists of $\phi_1({\bm r}_h)$ and $\phi^{'}_1({\bm r}_e)$, the hole and electron orbitals respectively, with contribution $\lambda_1$. The second most important pair consists of $\phi_2({\bm r}_h)$ and $\phi^{'}_2({\bm r}_e)$, with contribution $\lambda_2$ (typically $0.9<\lambda_1 +\lambda_2<1$). Using Gaussian’s built-in routines, we form an NTO wave $\Psi$ by pointwise multiplication, which amounts to evaluating the 2-body wavefunction $\Psi({\bm r}_e,{\bm r}_h)$ at the electron-hole relative coordinate of zero, giving as a function of the center of mass coordinate ${\bm X}$:
\begin{equation}
	\Psi({\bm X}) = \sqrt{\lambda_1}\phi_1({\bm X})\phi^{'}_1({\bm X}) + \sqrt{\lambda_2}\phi_2({\bm X})\phi^{'}_2({\bm X}).
\end{equation}

Plotting $\Psi({\bm X})$ for the 10 nm case in Fig.\,\ref{supp-NTO_wave}, we see dumbbell-shaped lobes of positive sign on the $A$ carbon atoms of the graphitic sidewall of the SWCNT and similar lobes of negative sign on the $B$ carbon atoms. The NTO wave’s sign alternates between positive and negative around a C$_6$ ring of the SWCNT sidewall, and along armchair or zigzag chains in the wall, forming helices in accord with the chirality. This ideal waveform is distorted to varying degrees in SWCNTs of different lengths but is still exhibited clearly. 

Some lengths of (6,5) SWCNTs exhibit states that are anomalous in terms of the NTO analysis and do not strictly follow this signature waveform. While we think that these states are the best candidates for the $E_{11}$ in their respective models, we exclude these anomalous states from the dataset used for fitting, as indicated in Fig.\,\ref{fig2}. The results of the NTO analysis are presented for all (6,5) SWCNT lengths in the Supplemental Materials. All selected peaks are highlighted in the absorption spectra shown in Fig.\,\ref{fig1}c and referred to as $E_{11}$ in this paper.

In Fig.\,\ref{fig2}, we plot the selected $E_{11}$ peak energies for pristine (6,5) SWCNTs as a function of the scaled length $x=L/L^{(6,5)}_0$, with the length $L$ of the SWCNT in nanometers and the length scale chosen as $L^{(6,5)}_0=0.25$ nm. The quantum confinement effect is clearly seen as $x\longrightarrow 1$ cu, which corresponds to the Clar cell used to generate the models, as described in the Supplemental Materials. We note that the exciton size implied by the onset of the substantial blueshift is on the order of 2 nm, with shorter lengths rapidly causing a change in $E_{11}$ greater than 2\%. This length is similar to previous theoretical estimations of the exciton Bohr radii in small-diameter SWCNTs\cite{Perebeinos2004,Spataru2004,Capaz2006}. It is also important to note that this is similar to an early measurement of the electron-hole correlation length\cite{Luer2009}, a measure of the exciton size. A more recent measurement of the correlation length places it closer to 13 nm\cite{Mann2016}, and using our best fit parameters we find that the percent change in $E_{11}$ between $x$ and $x-1$ for lengths greater than 13 nm is negligible ($<0.2\%$). In other words, the vast majority of blueshift occurs in SWCNTs with lengths $\leq 13$ nm.

As a finite SWCNT can be thought of as a type of quantum dot, it is instructive to consider a simple approach\cite{Brus1983,Brus1984} used to understand the confinement blueshift in the original colloidal quantum dots. These dots are spheres characterized by a radius $R$, and the $R$-dependence\footnote{The overall $R$-dependence in the third term of Eq.\,(\ref{BrusE}) goes as $R^{-1}$ as the dependence is averaged out within the brackets.} of the shift in the lowest excited state energy relative to the bulk value ($R\longrightarrow\infty$) can be approximately modeled as 
\begin{equation}
	\Delta E_{\rm QD}(R) = \frac{\pi^2\hbar^2}{2\mu R^2} - \frac{1.8 (e^2/4\pi\varepsilon_0)}{\varepsilon_2 R} + \frac{(e^2/4\pi\varepsilon_0)}{\varepsilon_2 R}\left\langle\sum^{\infty}_{l=1}a_l\left(\frac{r}{R}\right)^{2l}\right\rangle
	\label{BrusE}
\end{equation}
In this expression, $\mu$ is the reduced effective mass of the electron and hole, and the charges are placed in a dielectric sphere of constant relative permittivity $\varepsilon_2$ surrounded by another dielectric of constant $\varepsilon_1$. The coefficients in the sum follow $a_l=(\kappa-1)(l+1)/(\kappa l + l +1)$, with $\kappa=\varepsilon_2/\varepsilon_1$. The brackets denote an expectation value with respect to the first $S$-wave basis function $\psi \propto \sin(\pi r/R)/r$, where $r$ is the radial coordinate of either the hole or electron. The three contributions are a kinetic term, a Coulomb attraction of the electron-hole pair, and a polarization term due to a difference in the dielectric constants of the nanostructure and its surroundings. When $\varepsilon_2=\varepsilon_1$, all the $a_l=0$. 

\begin{figure}
	\includegraphics[scale=0.17]{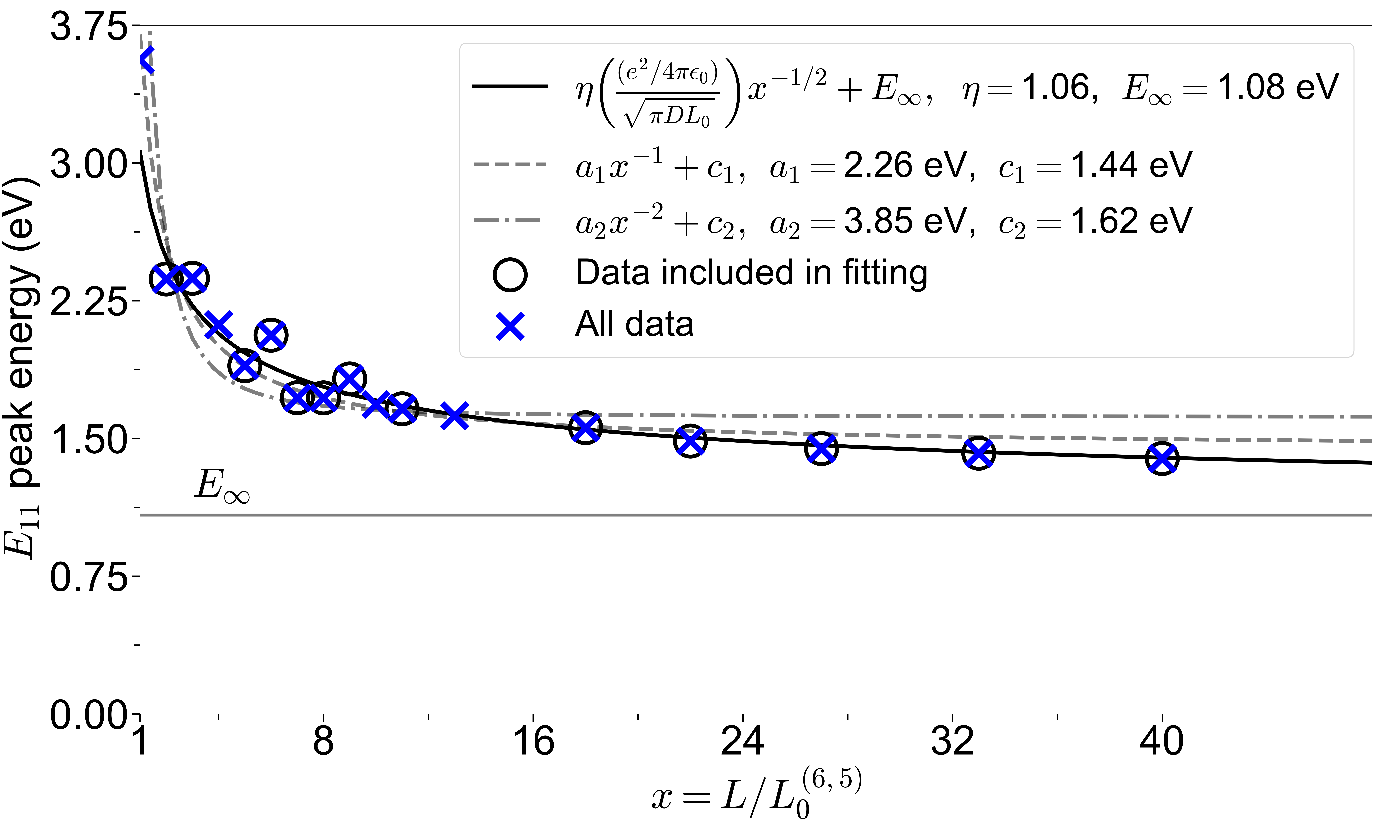}
	\caption{\textbf{TD-DFT calculations reveal an inverse square root dependence of the $E_{11}$ on nanotube length.} The length scale used is the length per Clar unit in nanometers for (6,5) SWCNTs, $L_0^{(6,5)}= 0.25$ nm. Possible fits from a 1D particle-in-a-box picture are plotted for comparison to illustrate the high quality of the $L^{-1/2}$ law. The $x=1$ cu case blueshifts higher than the trend, likely due to the geometry optimization, which for the nanobelt structures tends to twist adjacent C$_6$ rings \cite{Bachrach2010}.}
	\label{fig2}
\end{figure}
\begin{figure}
	\includegraphics[scale=0.08]{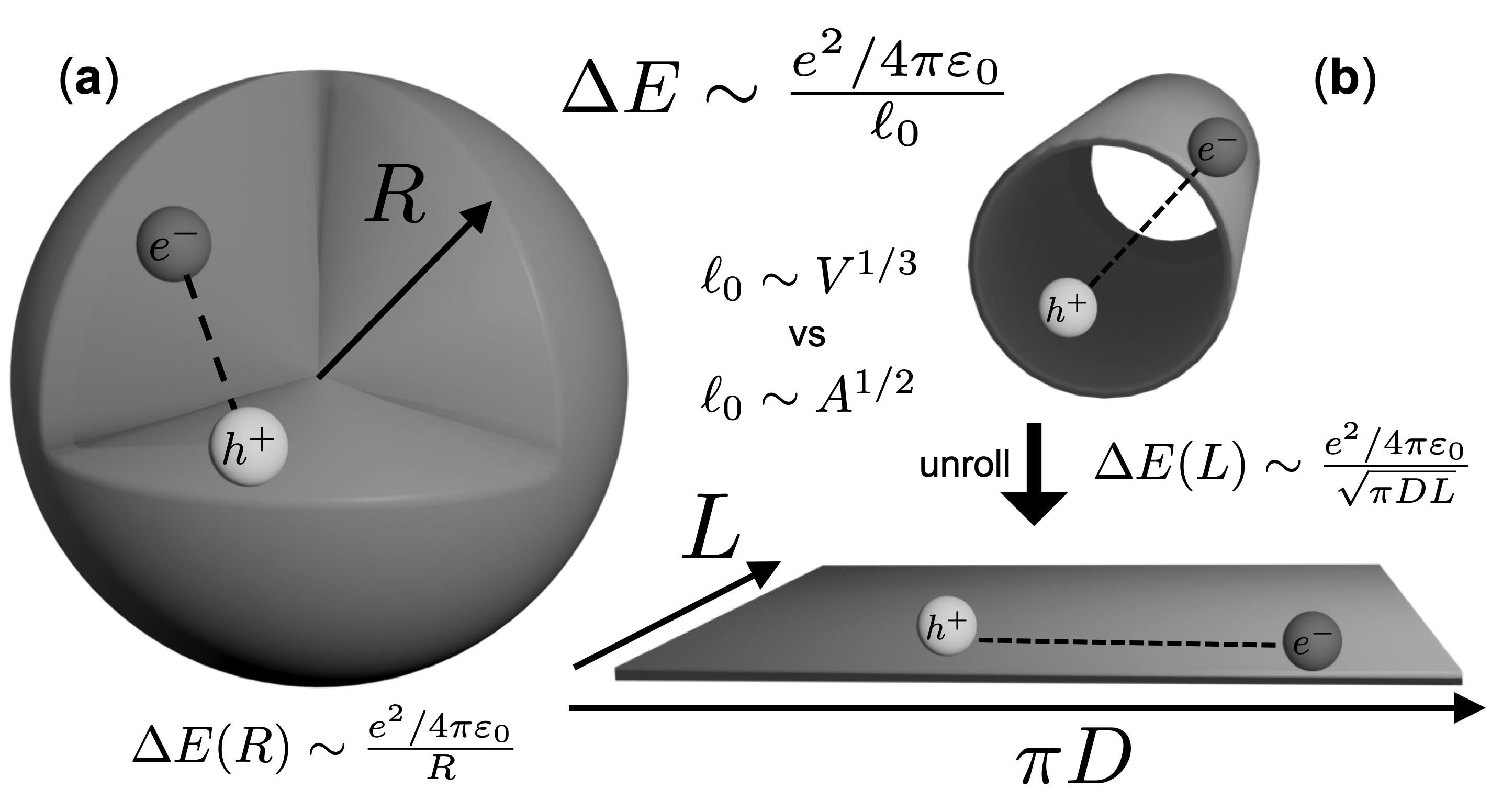}
	\caption{\textbf{Physical origin of the observed $L^{-1/2}$ dependence of the $E_{11}$.} An electron-hole pair in the space is defined by the boundaries of \textbf{(a)} the spherical structure and \textbf{(b)} the cylinder. Schematically, the charges in the quantum dot architecture exist in a volume $V$ determined by a single length scale, radius $R$, and in a SWCNT the charges exist in the surface area $A$ of the nanotube cylinder determined by the two independent length scales, chirality-dependent diameter $D$ and length $L$.}
	\label{fig3}
\end{figure}

Taking Eq.\,(\ref{BrusE}) as a heuristic, we can analyze the expected contributions to the confinement blueshifts in ultrashort SWCNTs. For the spherical quantum dots studied with this approach in\cite{Brus1983,Brus1984}, the kinetic energy term provides the dominant contribution for small $R$, and the confinement causes a blueshift that goes as $R^{-2}$, i.e. a particle-in-a-box type dependence. Making an analogy between spherical quantum dots, the size of which is controlled by the radius $R$, and finite SWCNTs, the size of which is controlled by the length $L$, we can let $R\rightarrow L$ in Eq.\,(\ref{BrusE}) and compare the resulting length dependence of the blueshift to the TD-DFT calculated values of $E_{11}$ in (6,5) SWCNTs, plotted in Fig.\,\ref{fig2}.
We see that the length dependence does not follow $L^{-2}$ scaling, as expected for a 1D particle-in-a-box. Neither does $L^{-1}$ capture the overall trend, nor a combination of the two. Both of these functions increase too rapidly as they approach zero and decay too rapidly to zero as $L$ increases. 

Using the scaled lengths $x$, we make a conjecture that the length dependence of the $E_{11}$ blueshift goes as $x^{-1/2}$. This scaling is plotted as a solid black line in Fig.\,\ref{fig2}, and we see that it fits the overall trend. We can understand this length dependence in the following way. The Coulomb interaction between electron and hole is strong in SWCNTs due to their quasi-1D nature, as reflected in the large exciton binding energies measured in these systems\cite{Wang2005}. A contribution akin to the Coulomb terms in Eq.\,(\ref{BrusE}) should provide the majority of the blueshift. 

Dimensionally, this major contribution to the energy shift should go as $\Delta E_{11}\sim (e^2/4\pi\varepsilon_0)/\ell_0$, for some length scale $\ell_0$ characteristic of the confining geometry. The geometry of a SWCNT containing the exciton is that of an atomically thin cylindrical shell with an area determined by two independent length scales, the length $L$ and (chirality-specific) diameter $D$, as $A=\pi DL$. Then setting $\ell_0=\sqrt{A}$, we have 
\begin{equation}
	\Delta E_{11} \sim \frac{e^2/4\pi\varepsilon_0}{\sqrt{\pi DL}},
	\label{geom_shift}
\end{equation}
yielding the inverse square root dependence of the $E_{11}$ blueshift on the SWCNT length for a fixed diameter. 

\begin{figure*}
	\centering
	\begin{minipage}{0.5\textwidth}
		\raggedright
		\includegraphics[scale=0.09]{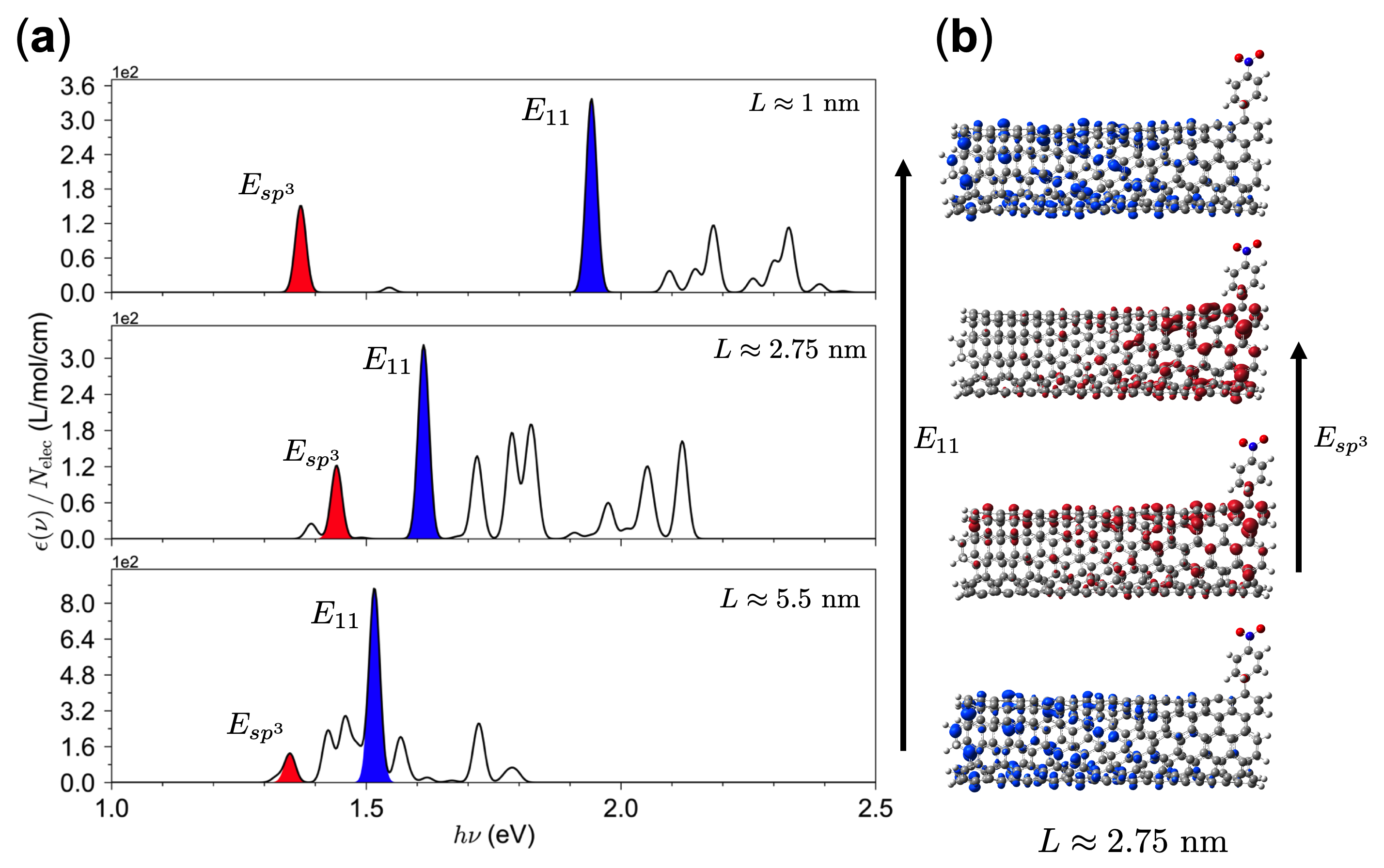}
		\label{fig4ab}
	\end{minipage}%
	\begin{minipage}{0.5\textwidth}
		\raggedleft
		\includegraphics[scale=0.09]{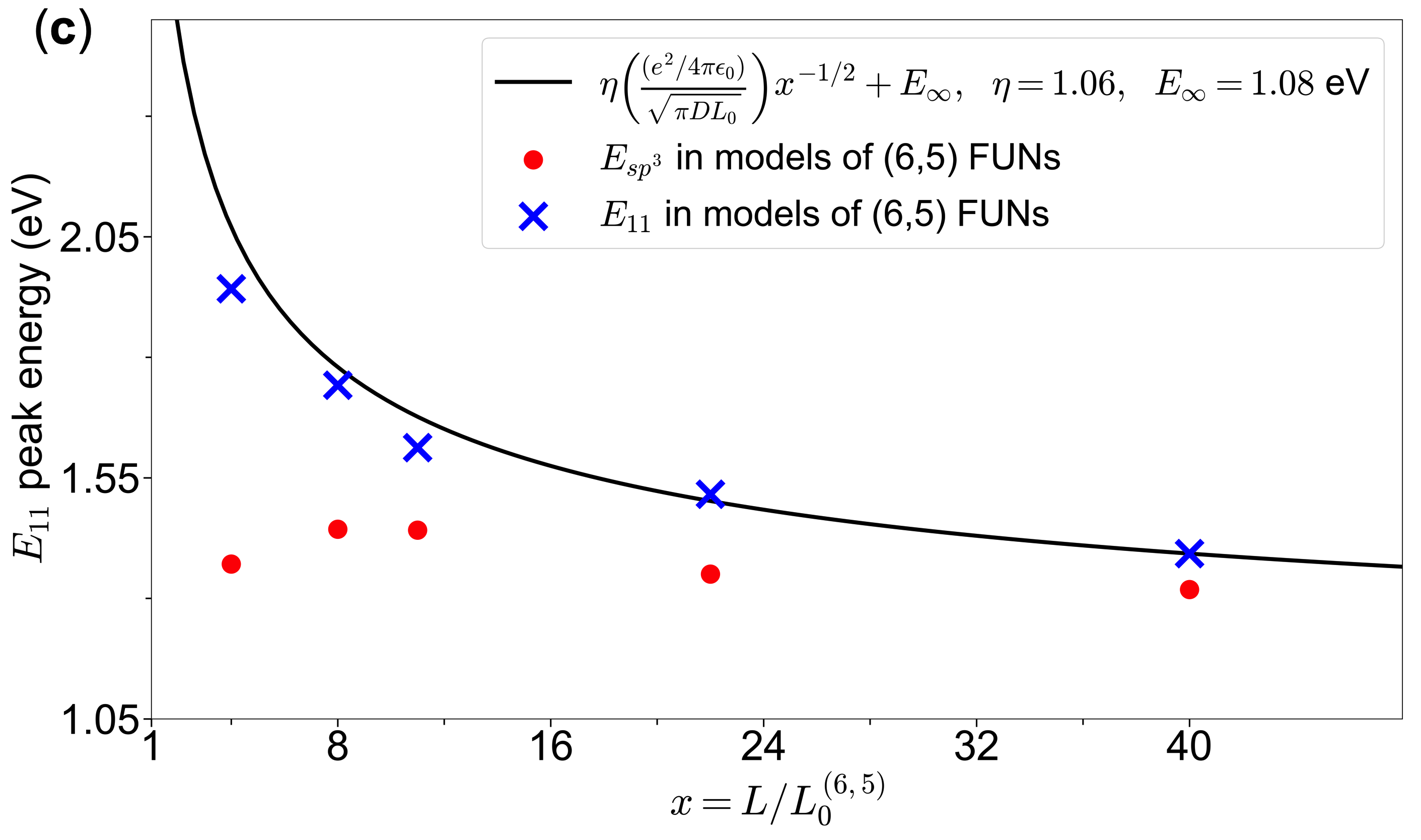}
		\label{fig4c}
	\end{minipage}
	
	\caption{\textbf{Quantum confinement manifested in fluorescent ultrashort nanotubes (FUNs).} \textbf{(a)} Calculated absorption spectra with highlighted defect-induced states $E_{sp^3}$, which are redshifted from the main $E_{11}$ peak associated with the pristine models. From top to bottom the lengths of the models are $x=4,11,22$ cu. \textbf{(b)} Squared NTOs describing the two highlighted transitions in $x=11$ cu (orbital coloring matches the highlighted peaks in Fig.\,\ref{fig4}a, color online), displayed as a transition from the hole densities $|\phi({\bm r}_h)|^2$ to the electron densities $|\phi^{'}({\bm r}_e)|^2$. The NTOs show that the $E_{sp^3}$ state is localized by the edge defect. \textbf{(c)} A scatterplot of $E_{11}$ and $E_{sp^3}$ values calculated in models of (6,5) FUNs, overlaid with the trendline for quantum confinement obtained by fitting $E_{11}$ values calculated in pristine (6,5) models. The $E_{11}$ values in the FUN models still follow the trend, and the defects induce lower-lying transitions that are less affected by the length of the SWCNT model, owing to the fact that these transitions involve orbitals that are localized to the defect.}
	\label{fig4}
\end{figure*}  

This dimensional argument, schematically depicted in Fig.\,\ref{fig3}, can also apply to the case of the spherical quantum dots. In this case, the exciton lives in a three-dimensional spherical volume $V$ characterized by the single length scale $R$. Here, $\ell_0 = V^{1/3}\sim R$, which recovers the proper $R$-scaling for the electrostatic terms in Eq.\,(\ref{BrusE}). Another way to think about the preceding argument is that if the nanostructure geometry is defined by $f$ independent length scales $L_1,L_2,...,L_f$, then each length scale should enter into the expression on equivalent footing, and a geometry-induced shift stemming from a Coulomb interaction should go as $\Delta E \sim (e^2/4\pi\varepsilon_0)\left(\prod^f_{k} L_k\right)^{-1/f}$. 

To understand how well the length dependence suggested by Eq.\,(\ref{geom_shift}) fits our calculated $E_{11}$ transitions, we can write the scaling law using the scaled lengths $x=L/L_0$:
\begin{equation}
	E_{11}(x) = \eta\left(\frac{e^2/4\pi\varepsilon_0}{\sqrt{\pi D L_0}}\right)x^{-1/2} + E_{\infty}
	\label{invroot}
\end{equation}
Here the free parameters are the infinite-length limit $E_{\infty}$ and the dimensionless pre-factor $\eta$. The dimensionful scale in the brackets $E_0 = (e^2/4\pi\varepsilon_0)/\sqrt{\pi D L_0}$ is close to 1.9 eV for (6,5) SWCNTs, using $\pi D=\sqrt{273}a_{cc}$ with a carbon-carbon bond length of $a_{cc}=1.44$ \AA \cite{CNT_book_saito_dressel} and our chosen length scale $L_0 = L^{(6,5)}_0$. From Fig.\,\ref{fig2}, the best fit gives $\eta=1.06$, indicating that the dimensional argument predicts the shift within a factor of order unity. The geometry of the SWCNT combined with the strong Coulomb interaction between the electron and hole leads to a length dependence in the $E_{11}$ of (6,5) SWCNTs distinct from that expected for a 1D particle-in-a-box.

Furthermore, we can apply the scaling law (\ref{invroot}) to the shifts in previously reported\cite{Eller2022} TD-DFT calculated excited states in ultrashort metallic (6,6) SWCNTs. These states are not the $E_{11}$ state in the (6,6) SWCNTs, but rather optical transitions allowed by the finite length of armchair nanotubes \cite{Zhou2004,Eller2022}. We refer to these transitions as $E_{\rm conf}$. The calculations previously reported\cite{Eller2022} are recreated here using a 3-21g basis set instead of a 6-31g, which does not change the qualitative trend. Additional models with $x=4$, 6, and 83 cu are included, which correspond to approximate lengths of 0.5, 0.75, and 10 nm using the length scale $L^{(6,6)}_0 \approx 0.125$ nm. The peak energies are plotted in Fig.\,\ref{supp-scaling_law_6-6_cnts} and we find that the inverse square root length scaling fits the $E_{\rm conf}$ energies well, producing the zero-frequency long-length limit\cite{Zhou2004} better than the inverse length or inverse square fits ($E_{\infty}=0.04$ eV), and with a pre-factor still of order unity ($\eta = 2.17$).

Finally, we connect this scaling law to models of FUNs. In Fig.\,\ref{fig4}a, we plot calculated absorption spectra of select FUN models ranging in length from 1 to 5.5 nm. By comparing with the corresponding pristine spectra, we identify the $E_{11}$ transition in the FUNs. Additionally, we find that the edge defects introduce redshifted transitions in a manner similar to previous TD-DFT calculations on finite SWCNTs with $sp^3$ defects\cite{Eller2022,Kilina2012}. These transitions, labeled $E_{sp^3}$, appear in different length FUNs. The NTO description of these excited states demonstates their localization at the edge defect, an example of which is shown in Fig.\,\ref{fig4}b for the $x=11$ cu FUN. The presence of redshifted, edge-localized excited states is reminiscent of transitions seen in PL spectra of FUNs\cite{Danne2018,Li2019}. 

Figure \ref{fig4}c shows the $E_{11}$ and $E_{sp^3}$ peak values for the FUNs plotted against the scaling law Eq.\,(\ref{invroot}) using the fit parameters extracted from the pristine (6,5) models. It is clear that the $E_{11}(x)$ trend persists in the FUNs, while the $E_{sp^3}$ state has little dependence on length due to its localized nature around the defect. There are intriguing prospects for systematic access to the ultrashort regime using the DICE method, which would enable an experimental test of the inverse square root length dependence of $E_{11}$ seen in our calculations here.

\section{Conclusion}
In conclusion, our calculations of the quantum confinement shift of the lowest exciton energy in ultrashort SWCNTs using the B3LYP approximation of TD-DFT reveal that the blueshift of the $E_{11}$ scales with length as $\Delta E_{11}\sim L^{-1/2}$. This length dependence departs from the expected 1D particle-in-a-box scaling. The onset of significant blueshift occurs when the length of the SWCNT approaches the exciton Bohr radius. We find this scaling in pristine ultrashort (6,5) and (6,6) SWCNTs, as well as in (6,5) models of FUNs that have an $sp^3$ defect pair installed near one edge. In the FUNs, we also find $sp^3$ defect states whose energies have minor dependence on the length, owing to the edge defect localized nature of these excited states. The appearance of the $L^{-1/2}$ scaling in both semiconducting and metallic chiralities, as well as in SWCNTs with $sp^3$ defects, suggests a broad applicability of the observed scaling to ultrashort SWCNT systems.   

\begin{acknowledgments}
	This material is based upon work supported in part by the National Science Foundation under award No. 1839165. The authors would like to thank Jacek K\l{}os, Jacob Fortner, Michael Winer, Andy Sheng and Alexandra Brozena for valuable discussions during the development of this work. We would also like to acknowledge the Alexander Family for their generous support through the Alexander Family Fellowship for students of the Chemical Physics Program at the University of Maryland. Furthermore, we acknowledge the high-performance computing team running the Zaratan Cluster at the University of Maryland, without which this research could not have been completed. 
\end{acknowledgments}

%\label{}
%\subsection{}
%\subsubsection{}

% If in two-column mode, this environment will change to single-column format so that long equations can be displayed. 
% Use only when necessary.
%\begin{widetext}
%$$\mbox{put long equation here}$$
%\end{widetext}

% Figures should be put into the text as floats. 
% Use the graphics or graphicx packages (distributed with LaTeX2e).
% See the LaTeX Graphics Companion by Michel Goosens, Sebastian Rahtz, and Frank Mittelbach for examples. 
%
% Here is an example of the general form of a figure:
% Fill in the caption in the braces of the \caption{} command. 
% Put the label that you will use with \ref{} command in the braces of the \label{} command.
%
% \begin{figure}
% \includegraphics{}%
% \caption{\label{}}%
% \end{figure}

% Tables may be be put in the text as floats.
% Here is an example of the general form of a table:
% Fill in the caption in the braces of the \caption{} command. Put the label
% that you will use with \ref{} command in the braces of the \label{} command.
% Insert the column specifiers (l, r, c, d, etc.) in the empty braces of the
% \begin{tabular}{} command.
%
% \begin{table}
% \caption{\label{} }
% \begin{tabular}{}
% \end{tabular}
% \end{table}

% If you have acknowledgments, this puts in the proper section head.
%\begin{acknowledgments}
% Put your acknowledgments here.
%\end{acknowledgments}

% Create the reference section using BibTeX:
\bibliography{references}

\end{document}

\title{Scaling Law of Quantum Confinement in Single-Walled Carbon Nanotubes - Supplemental Material}
%Can finite armchair nanotubes host organic color centers?
% A theoretical study

%
% comment out for anonymizing
\author{Benjamin Eller}
\affiliation{Institute for Physical Sciences and Technology, University of Maryland, College Park, MD 20742, USA}
\author{Charles W. Clark}
\affiliation{Joint Quantum Institute, National Institute of Standards and Technology and the University of Maryland, Gaithersburg, MD 20899, USA}
\author{YuHuang Wang}
\affiliation{Department of Chemistry and Biochemistry, University of Maryland, College Park, MD 20742, USA}

\date{\today}

\begin{abstract}
Here is presented the application of the inverse square root length scaling law to the lowest bright excited state transition in ultrashort (6,6) single-walled carbon nanotubes (SWCNTs), details of how the absorption spectra were calculated and plotted, visualizations of natural transition orbitals (NTOs) of the selected $E_{11}$ states of (6,5) SWCNTs, and atomic coordinates of representative SWCNT models, including a model of a fluorescent ultrashort nanotube (FUN). All models and orbitals were visualized using GaussView \cite{gv6}. 
\end{abstract}

\maketitle

\onecolumngrid
\section{The inverse square root scaling law extended to metallic SWCNTs}
\begin{figure}[!h]
	\includegraphics[scale=0.27]{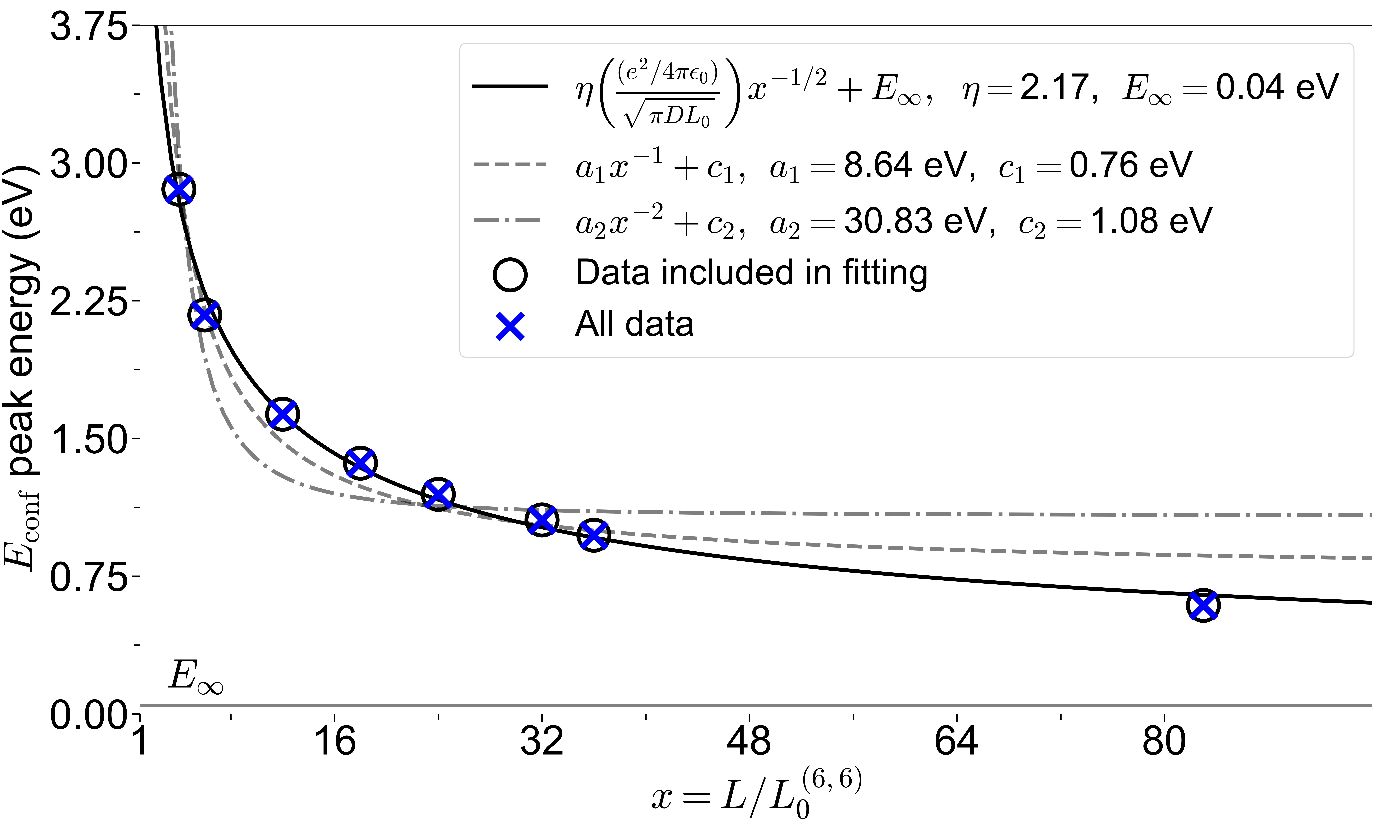}
	\caption{\textbf{The $L^{-1/2}$ scaling law extended to metallic SWCNTs.} The calculations of quantum confinement performed on ultrashort (6,6) SWCNTs following Eller, \textit{et al}\cite{Eller2022}. Note that the inverse square root scaling better captures the long-length metallic limit, whereas the $x^{-1}$ and $x^{-2}$ scalings do not.}
	\label{scaling_law_6-6_cnts}
\end{figure}

The length dependence of the quantum confinement shift in $E_{11}$ energy seen in (6,5) SWCNTs also applys to at least one class of optical transitions in ultrashort (6,6) SWCNTs. In the case of ultrashort (6,6) SWCNTs, this transition is not the $E_{11}$ exciton but a lower energy transition seen in previous TD-DFT calculations on armchair SWCNTs\cite{Zhou2004,Silva2016,Eller2022} that arises due to the confinement. The energy of this transition decays to zero as the length of the SWCNT goes to infinity, and we refer to this transition as $E_{\rm conf}$. In Fig.\,\ref{scaling_law_6-6_cnts} we plot the peak value of $E_{\rm conf}$, calculated with TD-DFT as described in Methods, as a function of length for pristine (6,6) SWCNTs. Using the scaled lengths $x=L/L^{(6,6)}_0$, we fit the inverse square root scaling law $E(x)=\eta E_0 x^{-1/2}+E_{\infty}$ to the $E_{\rm conf}$ values, with $L^{(6,6)}_0=0.125$ nm and $E_0 = (e^2/4\pi\varepsilon_0)/\sqrt{\pi D L^{(6,6)}_0}$ ($D$ being the diameter of the (6,6) SWCNT). Again, the fitting parameters are $\eta$ and $E_{\infty}$, and these are displayed in Fig.\,\ref{scaling_law_6-6_cnts}. We also fit the functions $f(x)=a_1 x^{-1}+c_1$ and $g(x)=a_2 x^{-2}+c_2$ for comparison, and we see that the inverse square root function predicts the correct limit as $x\longrightarrow\infty$, while the others do not.

\phantom{This text will be invisible}

\section{Calculation and plotting of absorption spectra}
The results of our TD-DFT calculations provide excitation energies $E_n$ in eV, and oscillator strengths $f_n$, for the first 15 excited states. Using this we calculate, following Gaussian's website\cite{gaussian_uvvis}, an absorption spectrum for an optimized molecular model with $N_{\rm elec}$ electrons as a function of energy in eV. More specifically, we calculate the molar attenuation coefficient per electron using:
\begin{equation}
\varepsilon(E) = N^{-1}_{\rm elec}\sum_{n} (hc/e)(1.3062974\times 10^8)(f_n/\sigma)\exp(-(E - E_n)^2/\sigma^2)
\label{epsilonE}
\end{equation}
Here, $\varepsilon$ is measured in ${\rm L}\,{\rm mol}^{-1}\,{\rm cm}^{-1}$, $\sigma,E$ and $E_n$ in eV, $h=6.626070040\times10^{-34}\,{\rm Js}$ is Planck's constant, $c=2.99792458\times10^{10}\,{\rm cm}\,{\rm s}^{-1}$ is the speed of light in vacuum, and $e=1.6021766208\times10^{-19}\,{\rm C}$ is the elementary charge. 

For the spectra calculated with Eq.\,(\ref{epsilonE}) and presented in Fig.\,\ref{main-fig1}c of the paper, a scaling procedure is used. This is done in order to properly display the spectra for different lengths of SWCNT in a single window, as the maximum oscillator strength varies for different models. In particular, the maximum oscillator strength per electron over the first 15 excited states grows roughly linearly with the length of the SWCNT. To fit all of the spectra into the window, we find the maximum value of Eq.\,(\ref{epsilonE}) across the whole dataset. If the length $L$ indexes the different spectra, we take $\varepsilon_{\rm MAX} = \max_{L}(\max_{E}\varepsilon_L(E))$, first finding the maximum value across all energies in the spectrum for each $L$ and then finding the maximum of this set of values. Then we define a scale factor for each spectrum as $\gamma_L = \varepsilon_{\rm MAX}/4\max_{E}\varepsilon_L(E)$. Then we plot $\gamma_L\varepsilon_L(E) + \delta_L$, where $\delta_L$ is a constant offset. 

\section{Computational Methods}
\begin{figure}[!h]%
	\includegraphics[scale=0.18]{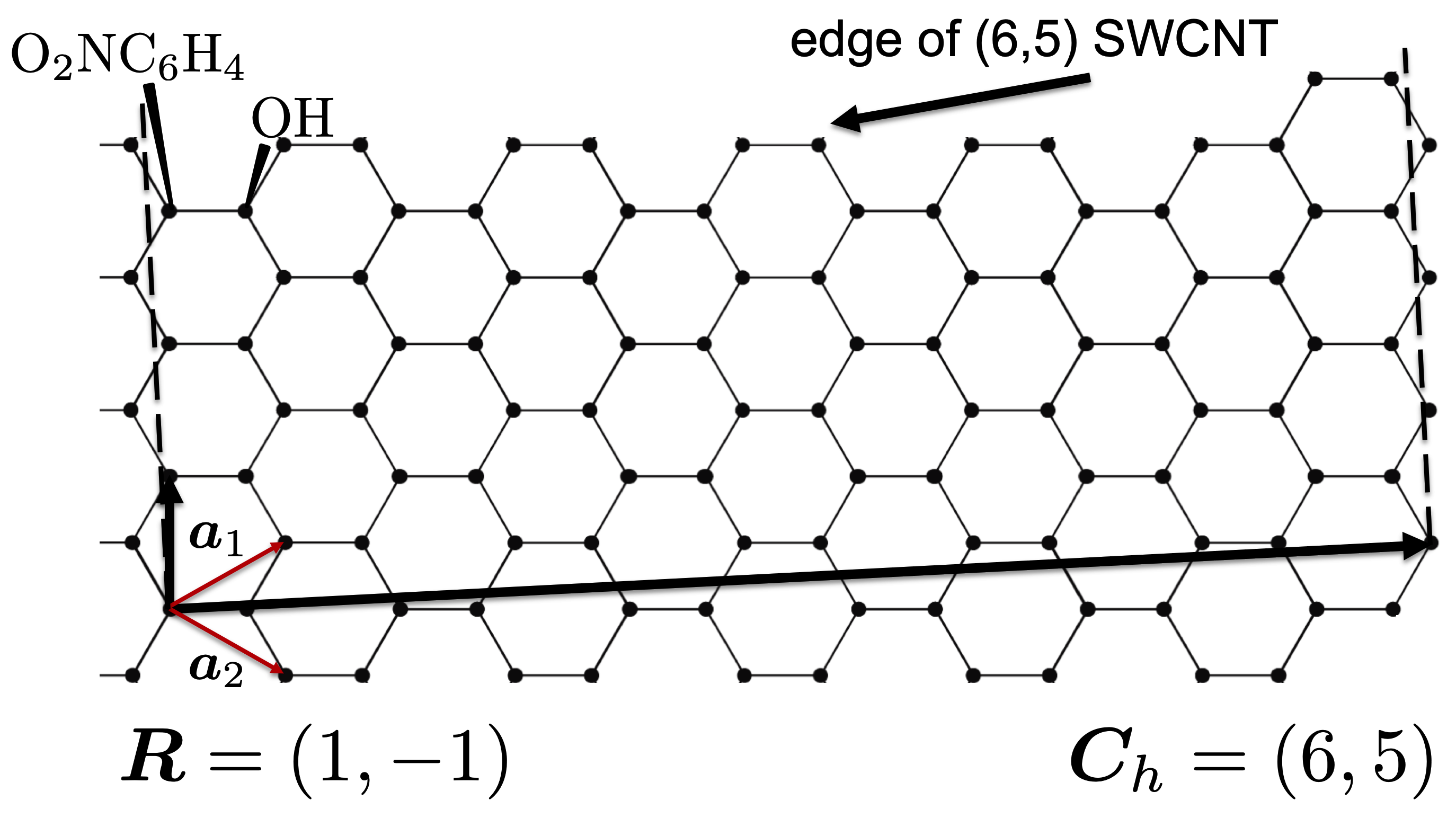}
	\caption{\textbf{Screw-axis transformation and $sp^3$ defect configuration.} Schematic for a (6,5) SWCNT showing the chiral vector ${\bm C}_h$ and the symmetry vector ${\bm R}$ written in the standard graphene basis ${\bm a}_1$ and ${\bm a}_2$. The periodic boundary condition in the circumferential (${\bm C}_h$) direction is indicated by the dashed lines at right angles to the ${\bm C}_h$ vector. The screw-axis transformation is a translation of the atoms by ${\bm R}$, which, coupled with the periodic boundary conditions, amounts to a translation along the nanotube axis combined with a rotation about that axis. Also shown is the $sp^3$ defect pair considered for the models of FUNs in this paper, and its configuration in relation to the edge of the finite SWCNT. The O$_2$NC$_6$H$_4$ refers to the aryl group, with the NO$_2$ in position 4 and the SWCNT in position 1 of the aromatic ring.}
	\label{model_schematic}
\end{figure}

All TD-DFT calculations in this paper were performed with Gaussian 16 computational chemical software\cite{g16} employing the B3LYP approximation\cite{B3LYP_1,B3LYP_2,B3LYP_3,B3LYP_4} for the exchange-correlation energy and a 3-21g Pople basis set\cite{Binkley1980}. This methodology is well suited for studying qualitative features of absorption spectra in $\pi$-conjugated carbon systems and is often used to study SWCNT systems\cite{Kilina2009,Kilina2012,Matsuda2010,Kwon2016,Gifford2019,Eller2022,Fortner2022,Qu2022}. The 3-21g basis set introduces minimal error but preserves the qualitative trends compared to a 6-31g basis set\cite{Kilina2009,Tretiak2007a}. The excited-state calculations were performed on geometry-optimized ground states calculated at the same level of theory. Unpaired electrons at the SWCNT edges were passivated with hydrogen atoms. We refer to the unit of length of the generated models as a Clar unit (cu), which maps to the number of applications of the screw-axis transformation defined by the SWCNT symmetry vector ${\bm R}$\cite{CNT_book_saito_dressel}, with 1 cu referring to the Clar cell (a nanobelt structure), 2 cu referring to the Clar cell plus atoms generated by 1 screw-axis operation on it (i.e. translation by ${\bm R}$), and so on. This construction is depicted schematically in Fig.\,\ref{model_schematic}. Thus, the molecular models have a chirality-specific length scale $L_0$, approximated in nm per cu, determined by measuring the length of the optimized models in nm and dividing by the number of cu. This length scale is related to the vector projection of the symmetry vector onto the axial direction, ${\bm R}\cdot{\bm T}/T$, where ${\bm T}$ is the translation vector of the SWCNT, with some variation due to relaxation of the ground-state geometries away from the ideal structure. For the (6,5) and (6,6) chiralities studied in this work the length scales are $L^{(6,5)}_0\approx 0.25$ nm and $L^{(6,6)}_0\approx 0.125$ nm. 

\section{Natural transition orbitals}
\label{one}
\begin{figure}[!h]%
	\includegraphics[scale=0.11]{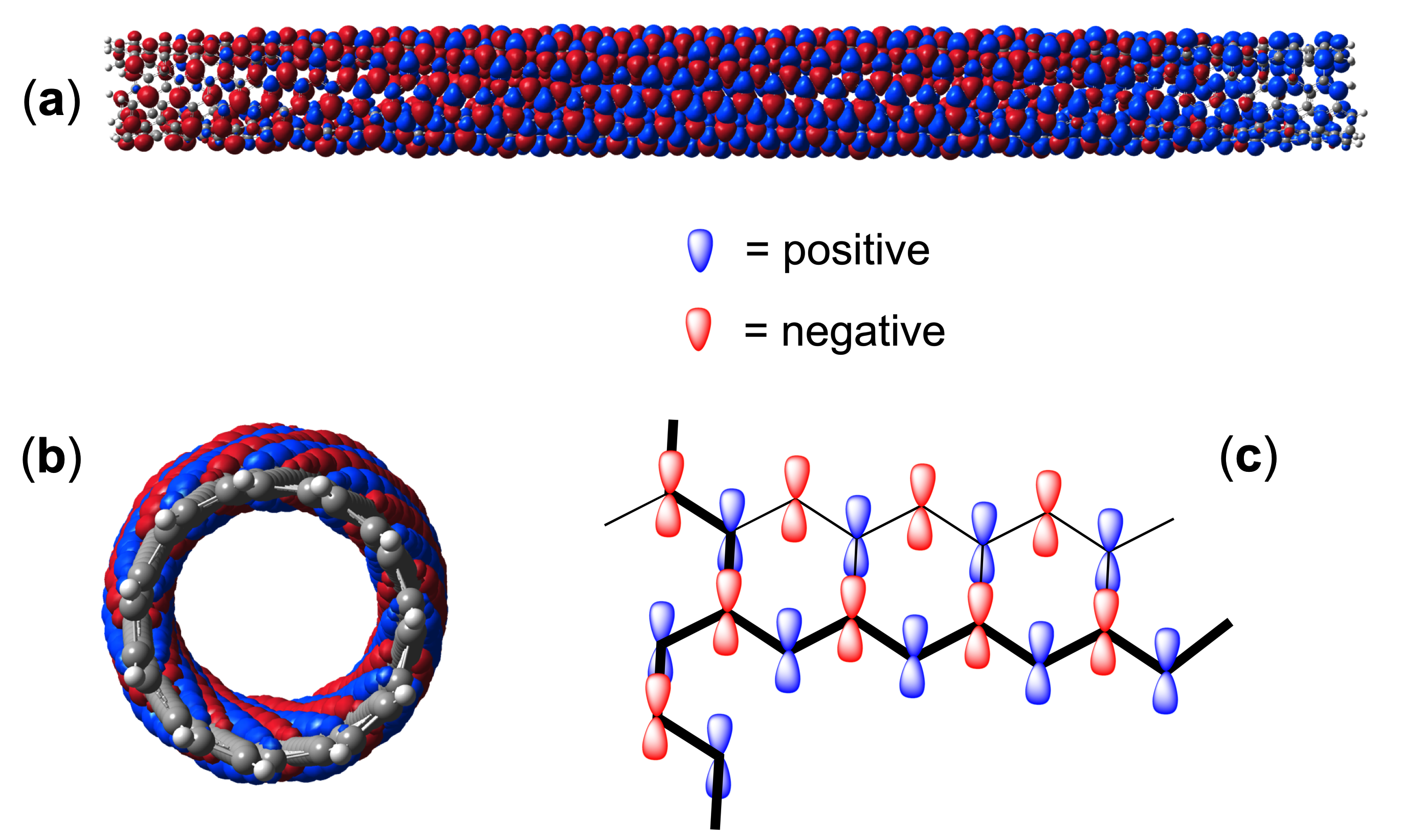}
	\caption{\textbf{NTO waves of the $E_{11}$ transition (color online).} \textbf{(a)} Side view of a 10 nm (6,5) SWCNT showing the NTO wave for the bright $E_{11}$ state, as well as \textbf{(b)} the axial view, showing the 11 periods of oscillation around the circumferential direction, corresponding to 11 edge C atoms in the (6,5). \textbf{(c)} Considering a subset of the SWCNT sidewall, we see that the waveform $\Psi({\bm X})$ for the $E_{11}$ in the 10 nm (6,5) displays an alternating sign around a C$_6$ ring face, with $A$ carbons having an antinode of positive sign (blue) and $B$ carbons an antinode of negative sign (red). The sign alternates moving along zigzag and armchair directions, forming helices in a manner dictated by the chirality.}
	\label{NTO_wave}
\end{figure}

In this section we provide the NTO waves for the designated $E_{11}$ states in (6,5) SWCNTs of different lengths calculated according to 
\begin{equation}
	\Psi({\bm X}) = \sqrt{\lambda_1}\phi_1({\bm X})\phi_1^{'}({\bm X}) + \sqrt{\lambda_2}\phi_2({\bm X})\phi_2^{'}({\bm X}).
\end{equation}
Here we use $\phi_1$ and $\phi^{'}_1$ to denote the hole and electron NTOs respectively, which share the largest singular value $\lambda_1$ (relative contribution to the excited state) found in the NTO transformation of the excited state \cite{Martin2003}, performed with Gaussian v16 software \cite{g16}. The NTO pair with the second largest singular value, $\lambda_2$, is denoted as $\phi_2$ and $\phi^{'}_2$, and the NTO wave is formed as the linear combination of these electron-hole pairs. Since the products are taken pointwise in Gaussian, the coordinate ${\bm X}$ used for plotting is the center of mass of the electron and hole. All isosurfaces are plotted at isovalue = 1e-5, and the color coding (color online) is such that blue represents a positive value of the NTO wave and red represents a negative value, as indicated in Fig.\,\ref{NTO_wave}.

\begin{figure}[!h]%
	\includegraphics[scale=0.18]{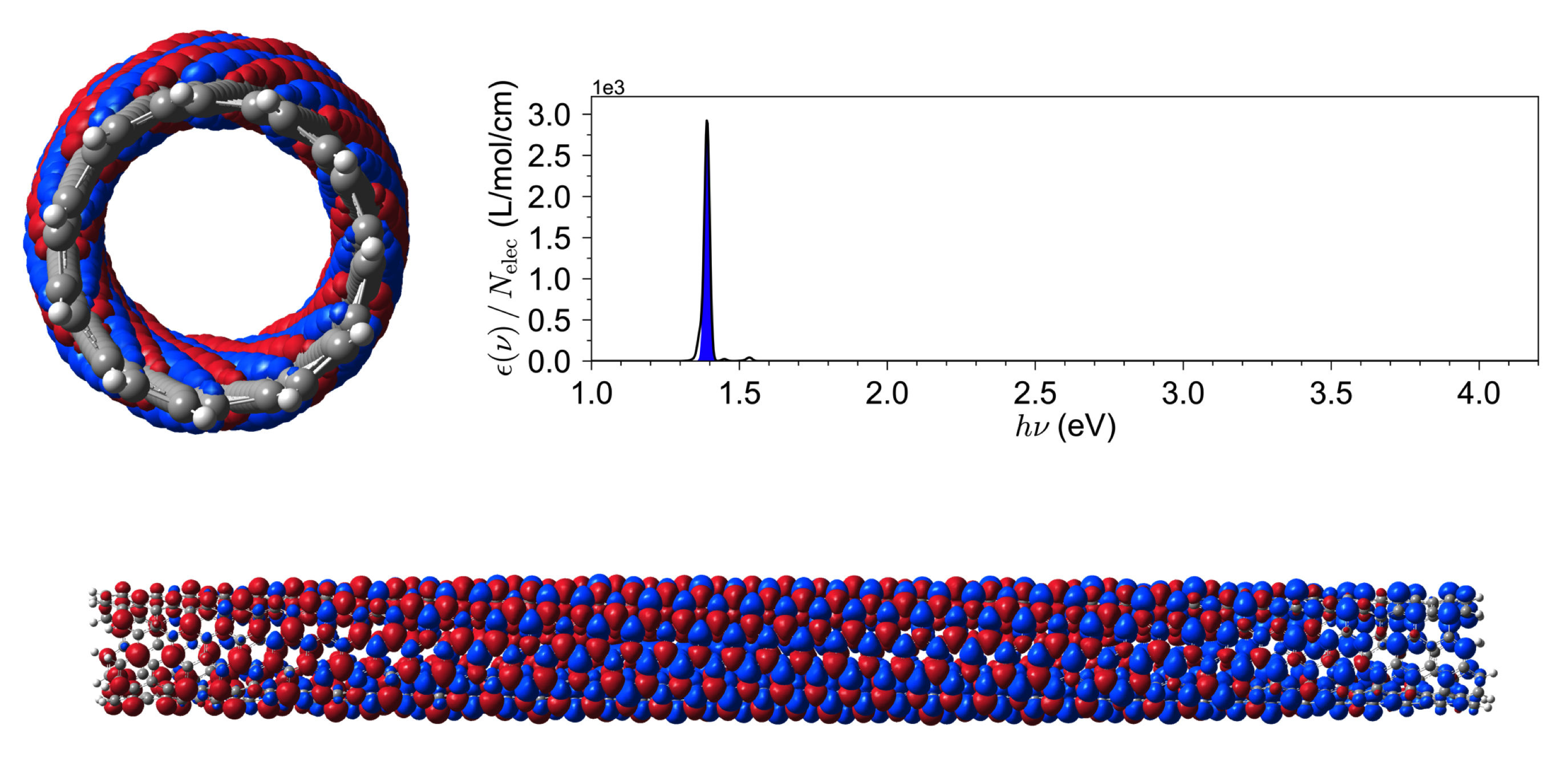}
	\caption{NTO wave of the $E_{11}$ transition, highlighted in blue in the spectrum, for a finite, pristine (6,5) SWCNT with a length of $x=40$ cu ($\approx$ 10.0 nm).}
	\label{NTO_wave_40cu}
\end{figure}

\begin{figure}[!h]%
	\includegraphics[scale=0.18]{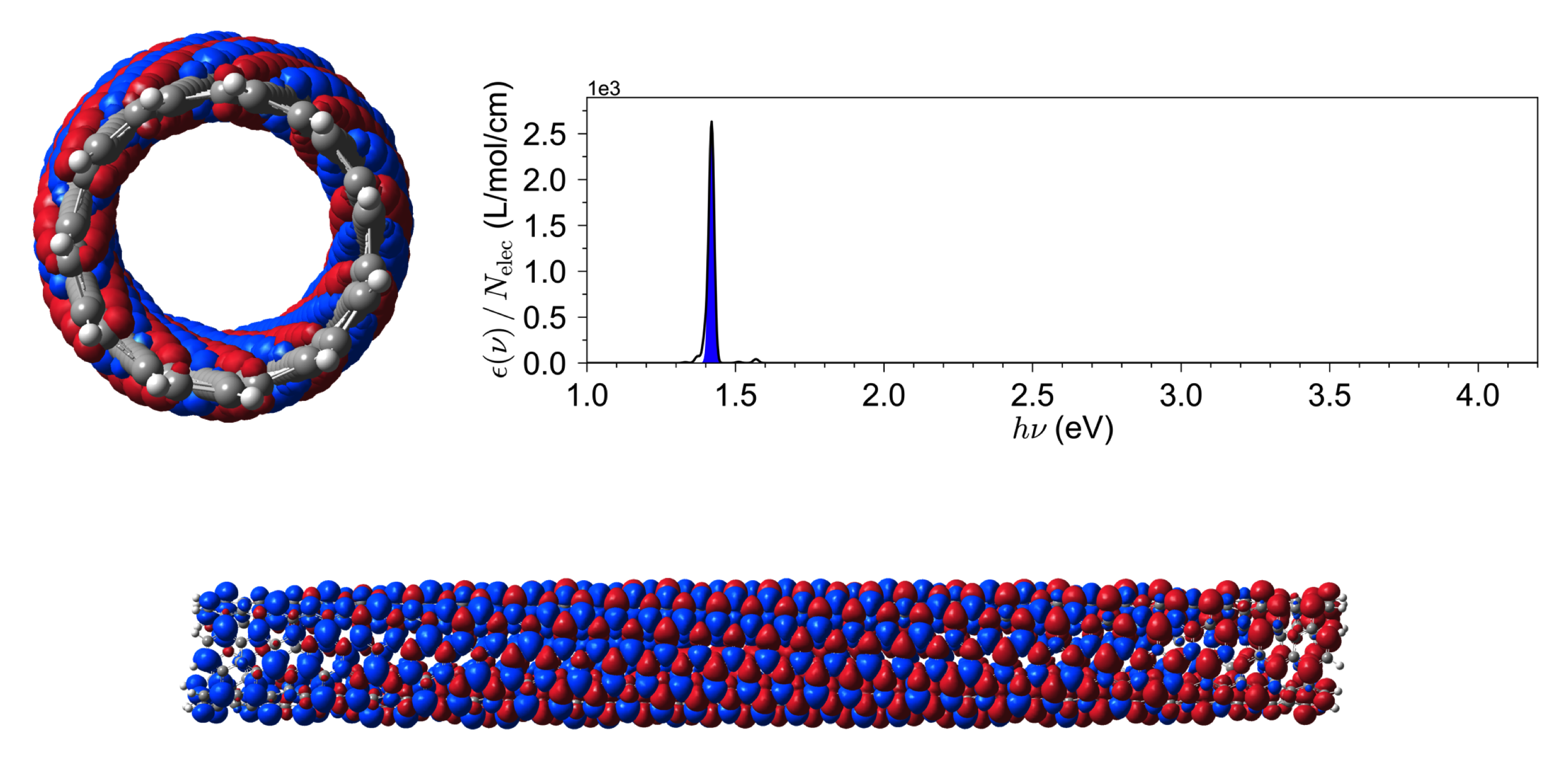}
	\caption{NTO wave of the $E_{11}$ transition, highlighted in blue in the spectrum, for a finite, pristine (6,5) SWCNT with a length of $x=33$ cu ($\approx$ 8.25 nm).}
	\label{NTO_wave_33cu}
\end{figure}
\newpage

\phantom{This text will be invisible}
\begin{figure}[!h]%
	\includegraphics[scale=0.18]{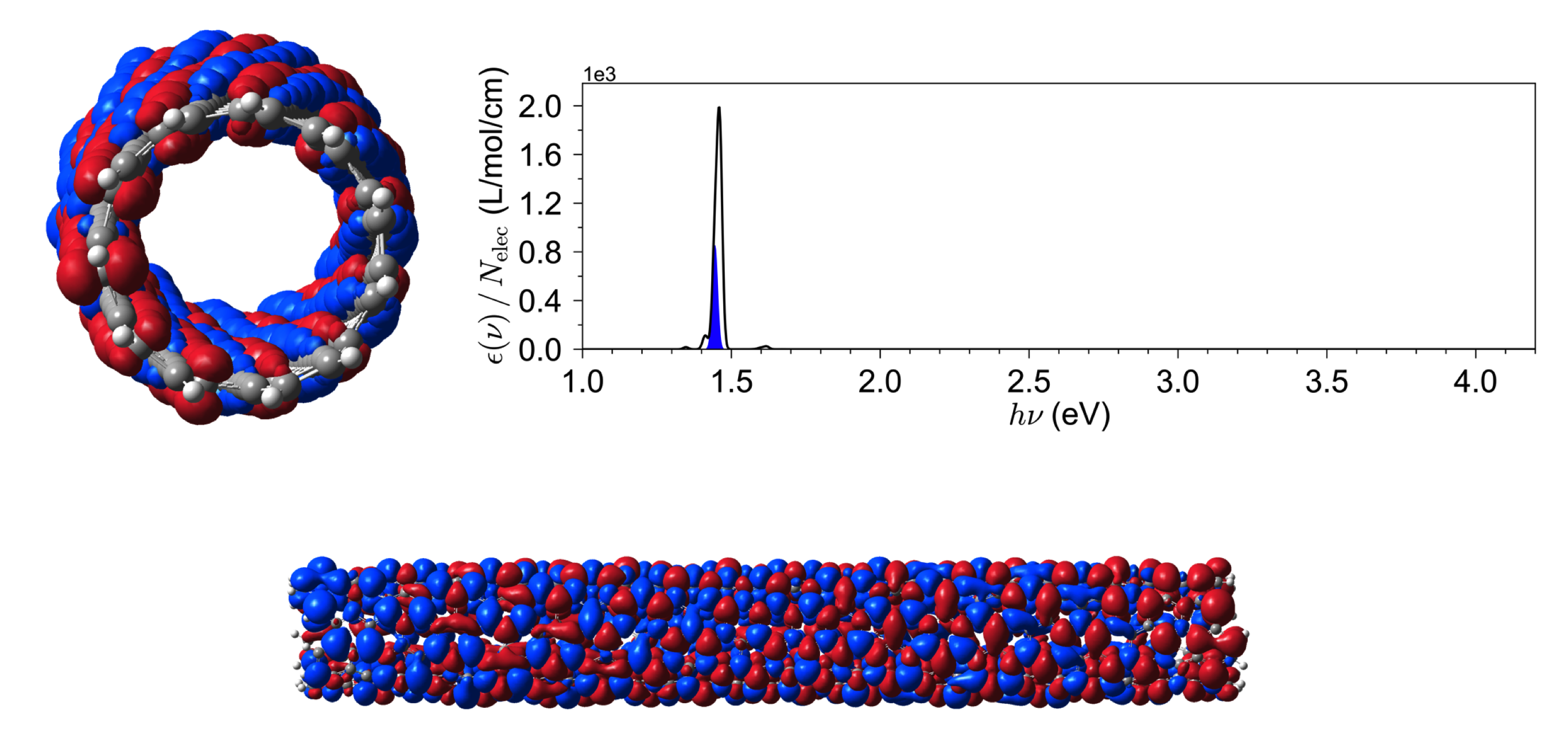}
	\caption{NTO wave of the $E_{11}$ transition, highlighted in blue in the spectrum, for a finite, pristine (6,5) SWCNT with a length of $x=27$ cu ($\approx$ 6.75 nm).}
	\label{NTO_wave_27cu}
\end{figure}

\begin{figure}[!h]%
	\includegraphics[scale=0.18]{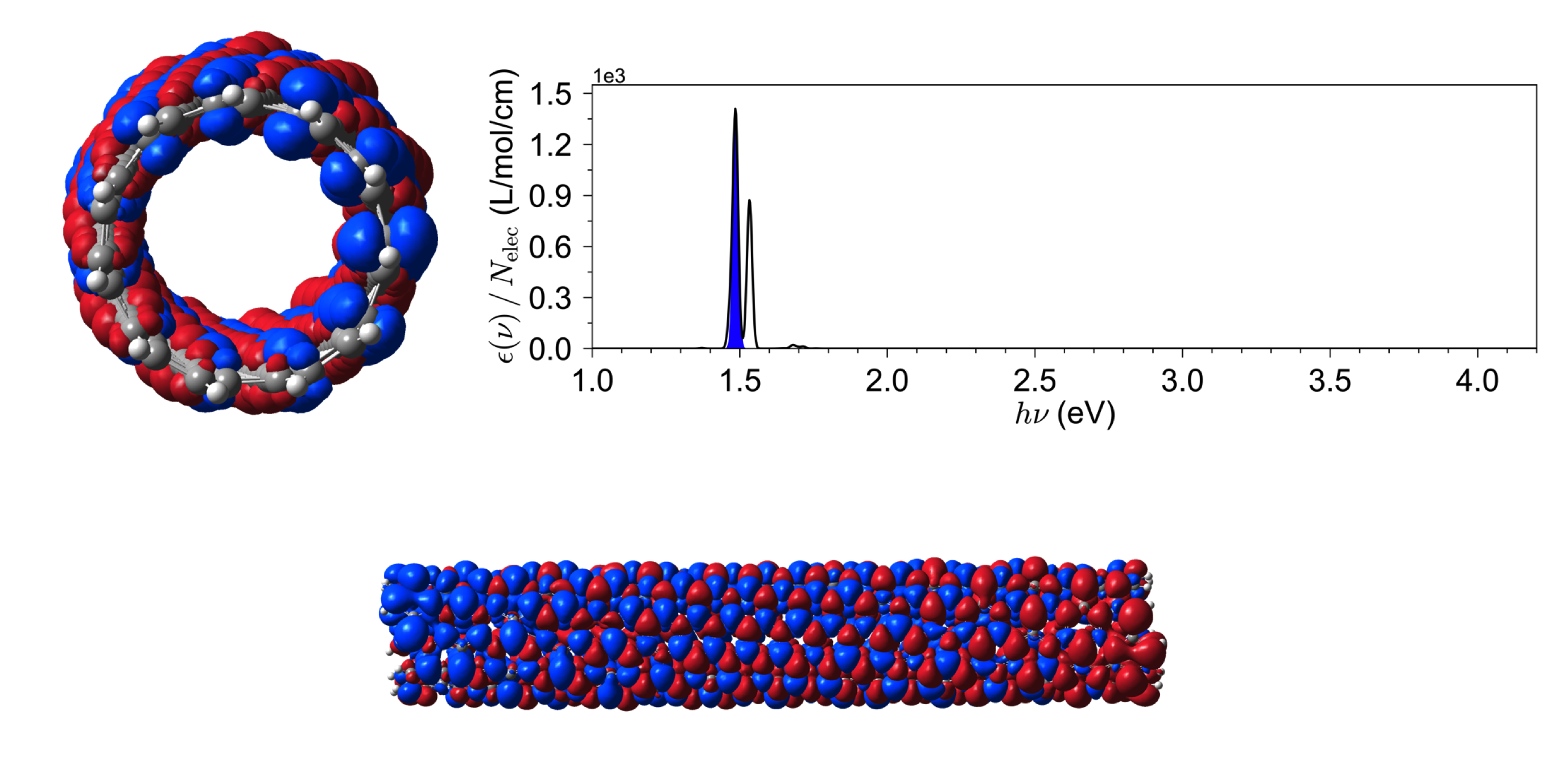}
	\caption{NTO wave of the $E_{11}$ transition, highlighted in blue in the spectrum, for a finite, pristine (6,5) SWCNT with a length of $x=22$ cu ($\approx$ 5.5 nm).}
	\label{NTO_wave_22cu}
\end{figure}
\newpage

\phantom{This text will be invisible}
\begin{figure}[!h]%
	\includegraphics[scale=0.18]{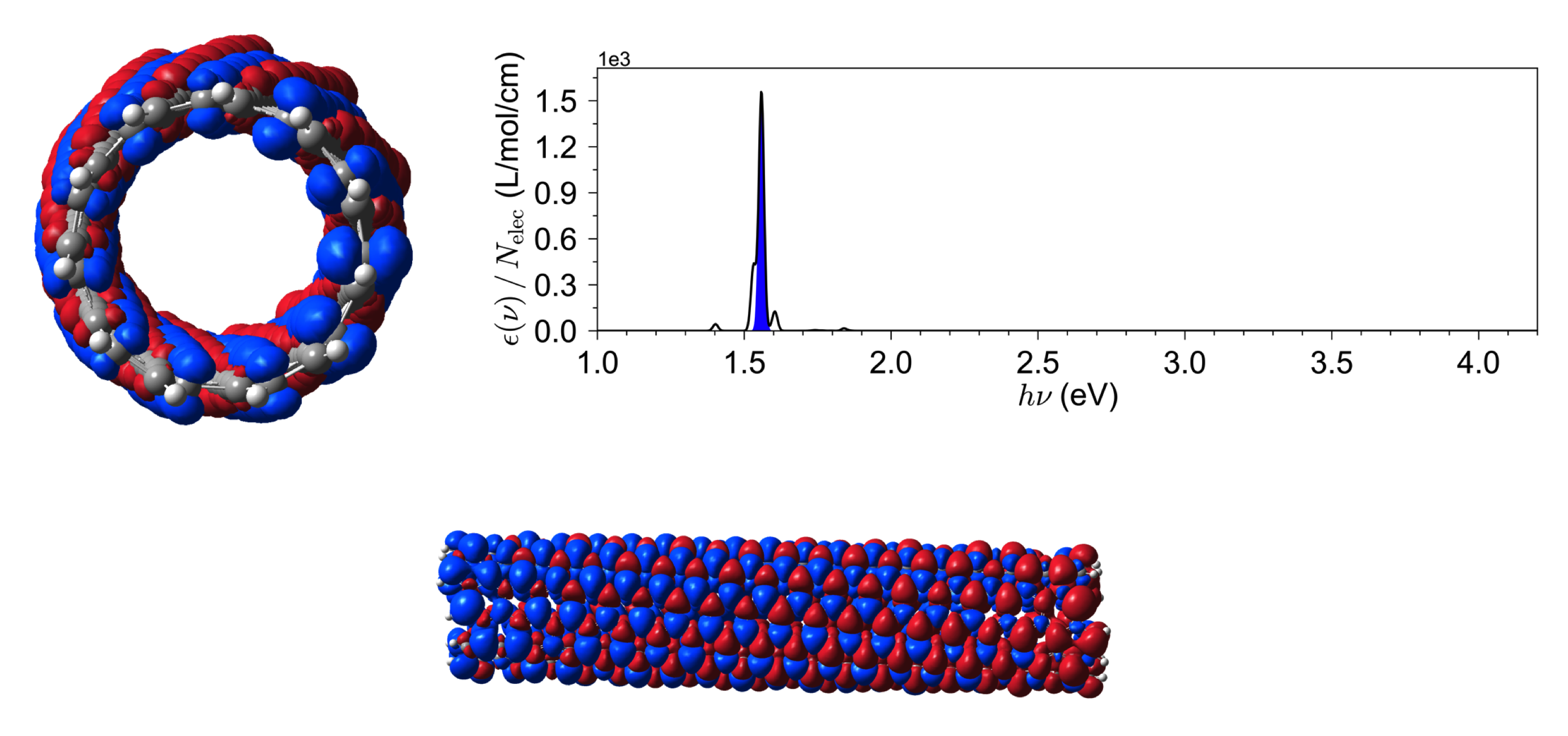}
	\caption{NTO wave of the $E_{11}$ transition, highlighted in blue in the spectrum, for a finite, pristine (6,5) SWCNT with a length of $x=18$ cu ($\approx$ 4.5 nm).}
	\label{NTO_wave_18cu}
\end{figure}

\begin{figure}[!h]%
	\includegraphics[scale=0.18]{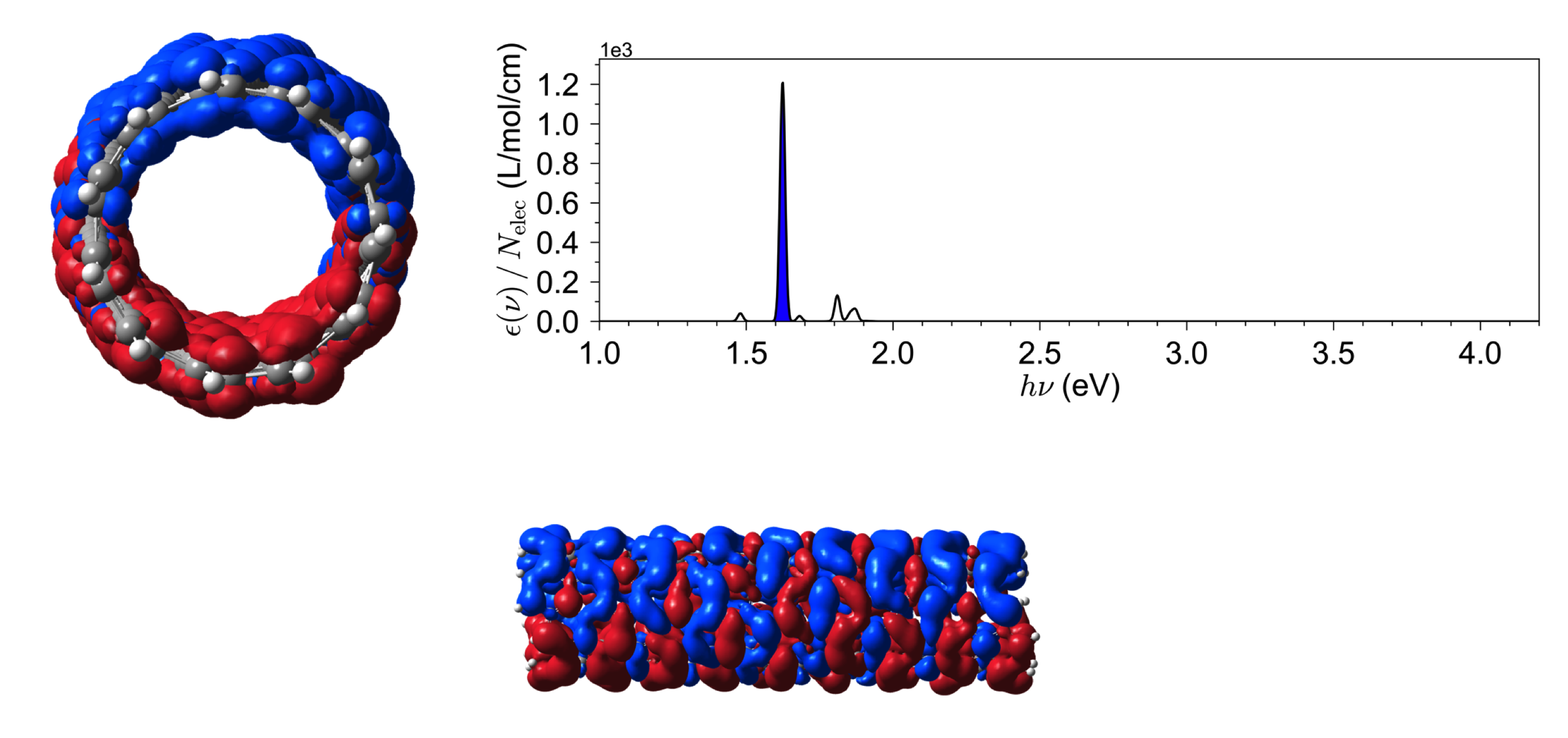}
	\caption{(Anomalous) NTO wave of the $E_{11}$ transition, highlighted in blue in the spectrum, for a finite, pristine (6,5) SWCNT with a length of $x=13$ cu ($\approx$ 3.25 nm).}
	\label{NTO_wave_13cu}
\end{figure}
\newpage

\phantom{This text will be invisible}
\begin{figure}[!h]%
	\includegraphics[scale=0.18]{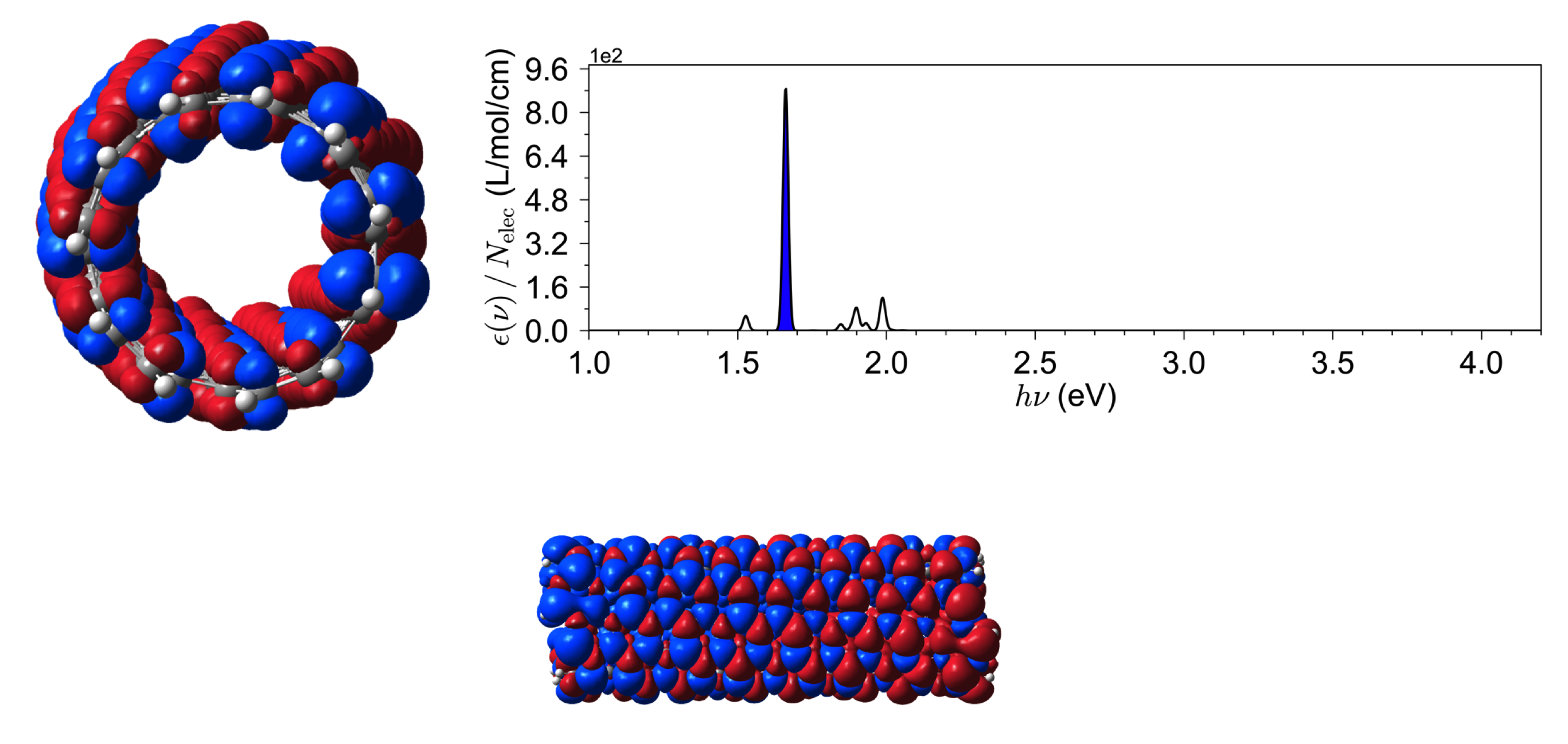}
	\caption{NTO wave of the $E_{11}$ transition, highlighted in blue in the spectrum, for a finite, pristine (6,5) SWCNT with a length of $x=11$ cu ($\approx$ 2.75 nm).}
	\label{NTO_wave_11cu}
\end{figure}

\begin{figure}[!h]%
	\includegraphics[scale=0.18]{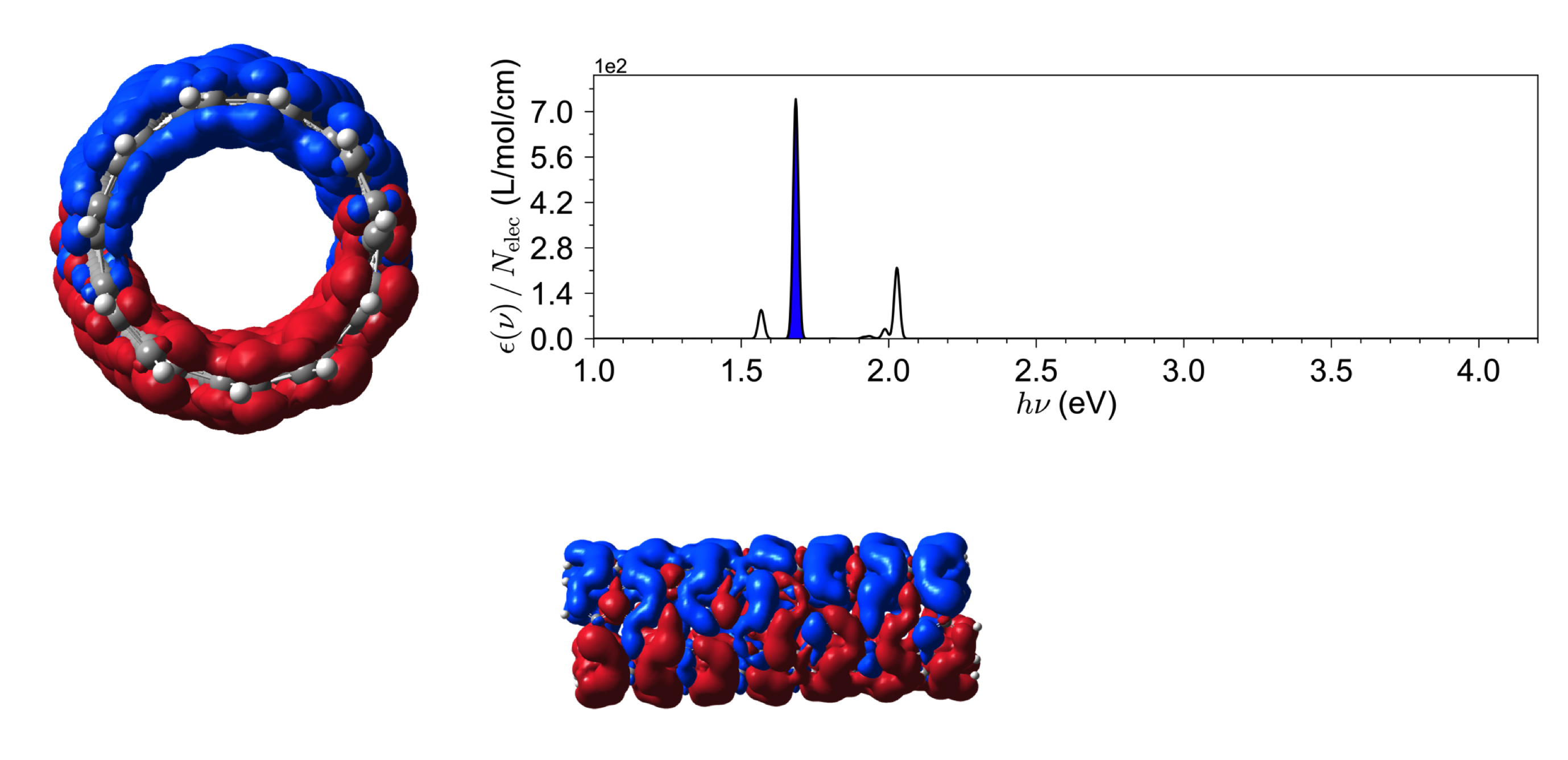}
	\caption{(Anomalous) NTO wave of the $E_{11}$ transition, highlighted in blue in the spectrum, for a finite, pristine (6,5) SWCNT with a length of $x=10$ cu ($\approx$ 2.5 nm).}
	\label{NTO_wave_10cu}
\end{figure}
\newpage

\phantom{This text will be invisible}
\begin{figure}[!h]%
	\includegraphics[scale=0.18]{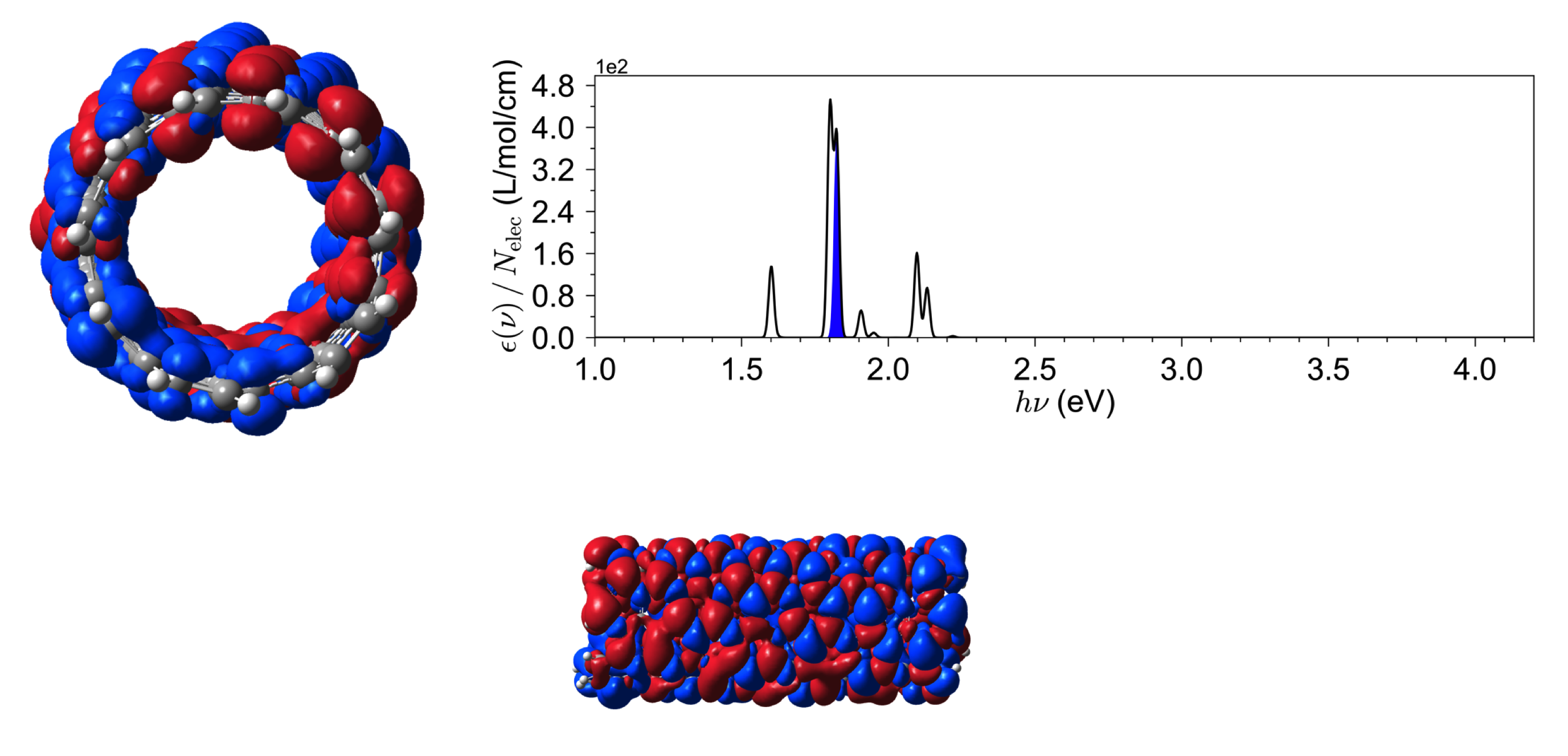}
	\caption{NTO wave of the $E_{11}$ transition, highlighted in blue in the spectrum, for a finite, pristine (6,5) SWCNT with a length of $x=9$ cu ($\approx$ 2.25 nm).}
	\label{NTO_wave_09cu}
\end{figure}

\begin{figure}[!h]%
	\includegraphics[scale=0.18]{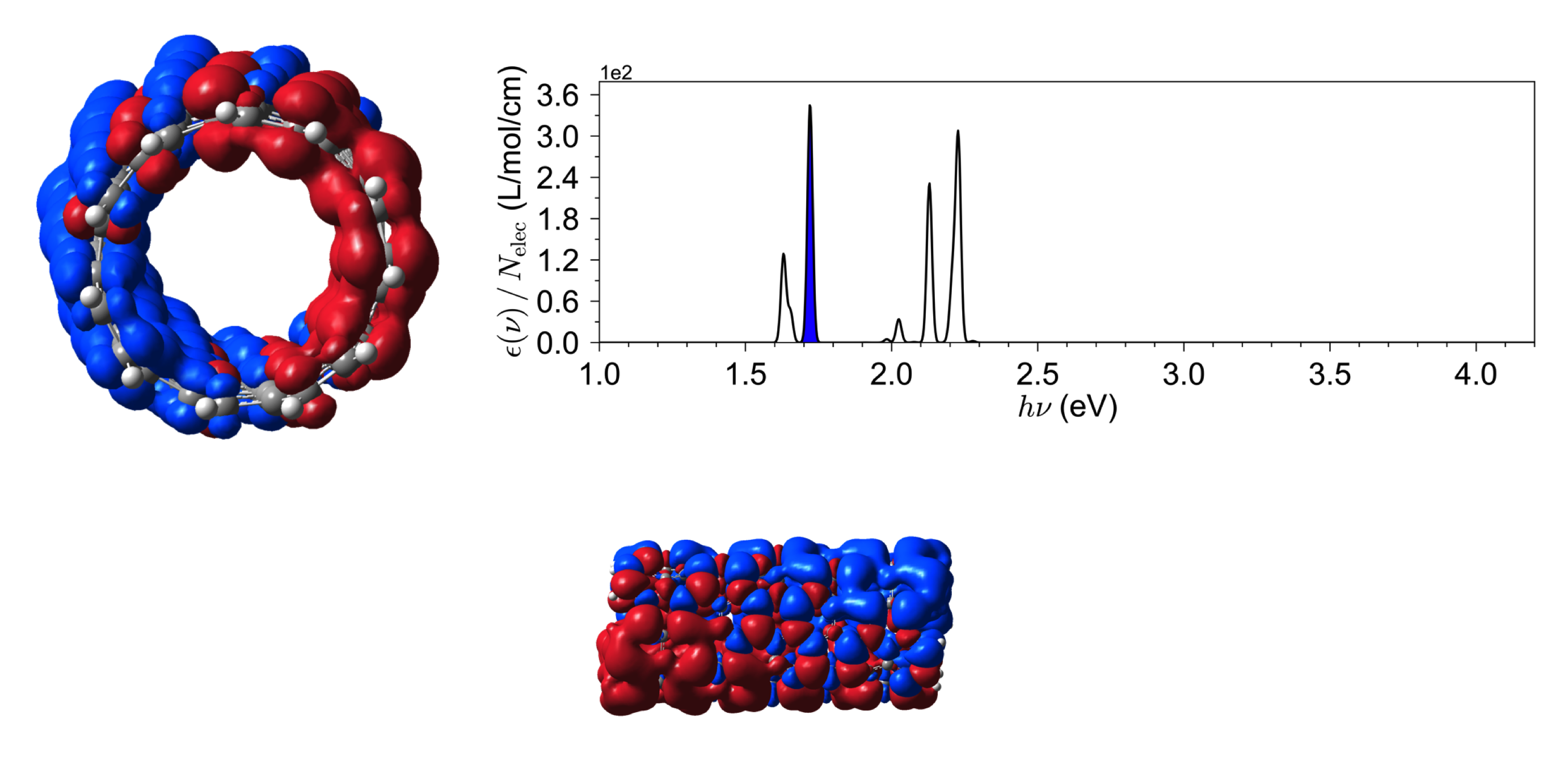}
	\caption{NTO wave of the $E_{11}$ transition, highlighted in blue in the spectrum, for a finite, pristine (6,5) SWCNT with a length of $x=8$ cu ($\approx$ 2.0 nm).}
	\label{NTO_wave_08cu}
\end{figure}
\newpage

\phantom{This text will be invisible}
\begin{figure}[!h]%
	\includegraphics[scale=0.18]{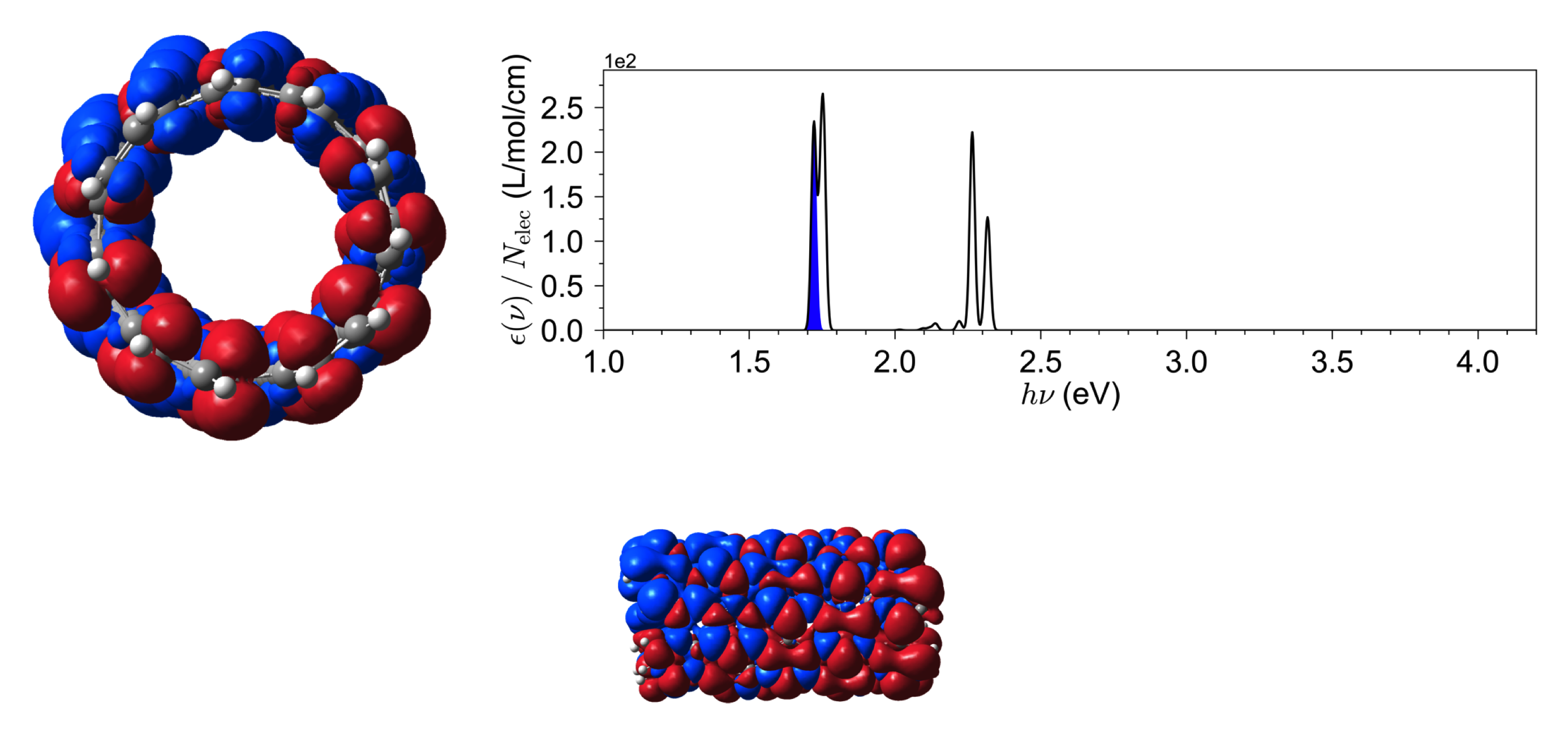}
	\caption{NTO wave of the $E_{11}$ transition, highlighted in blue in the spectrum, for a finite, pristine (6,5) SWCNT with a length of $x=7$ cu ($\approx$ 1.75 nm).}
	\label{NTO_wave_07cu}
\end{figure}

\begin{figure}[!h]%
	\includegraphics[scale=0.18]{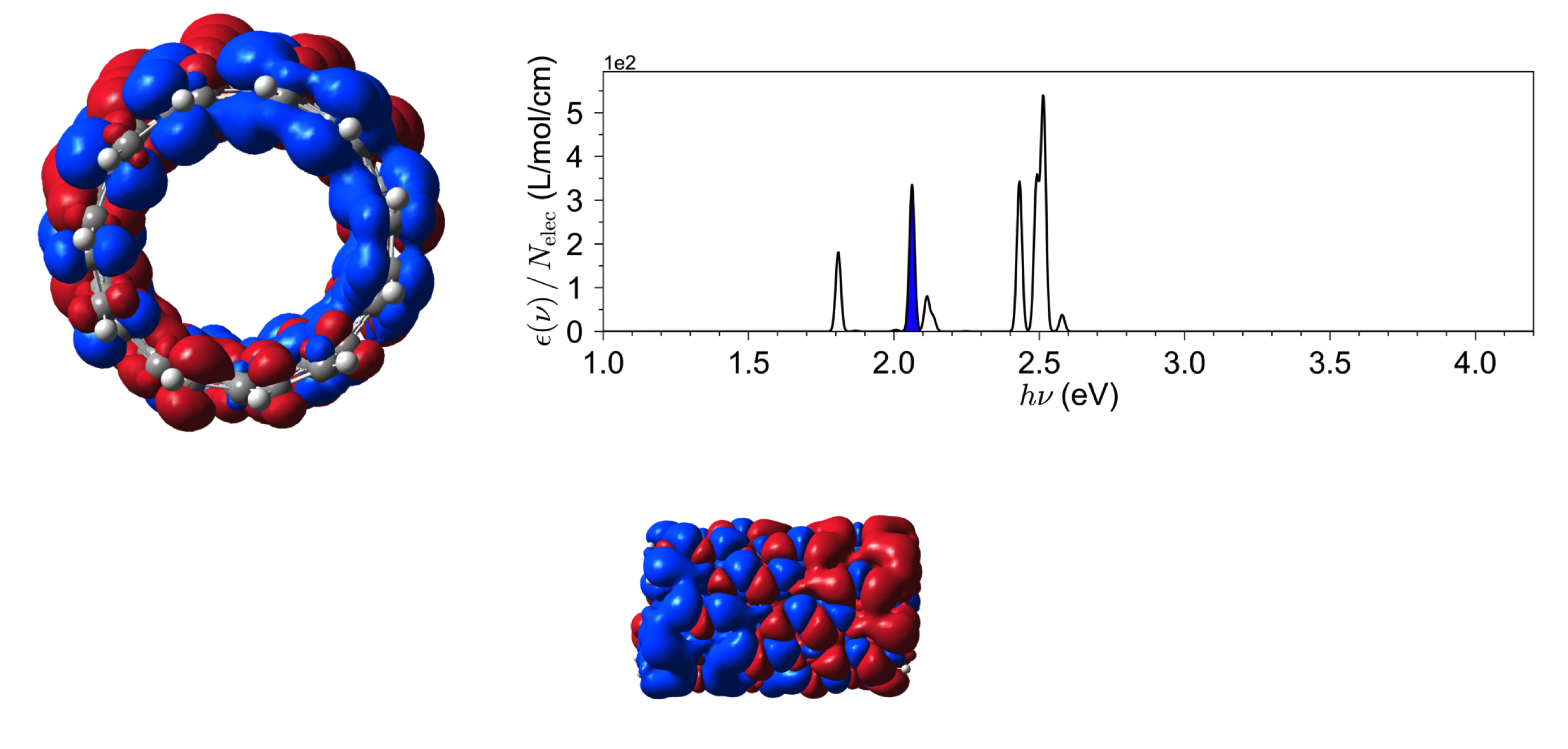}
	\caption{NTO wave of the $E_{11}$ transition, highlighted in blue in the spectrum, for a finite, pristine (6,5) SWCNT with a length of $x=6$ cu ($\approx$ 1.5 nm).}
	\label{NTO_wave_06cu}
\end{figure}
\newpage

\phantom{This text will be invisible}
\begin{figure}[!h]%
	\includegraphics[scale=0.18]{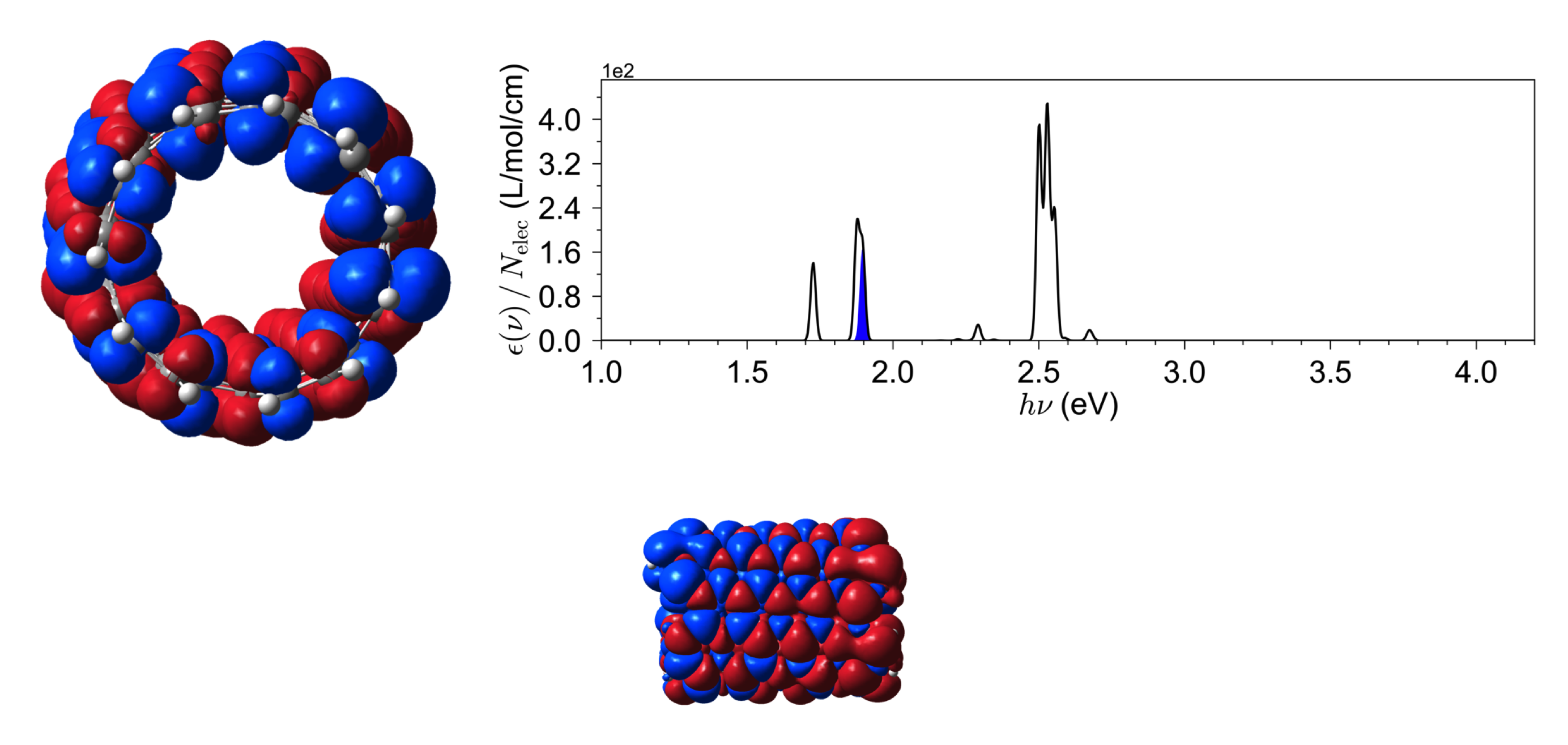}
	\caption{NTO wave of the $E_{11}$ transition, highlighted in blue in the spectrum, for a finite, pristine (6,5) SWCNT with a length of $x=5$ cu ($\approx$ 1.25 nm).}
	\label{NTO_wave_05cu}
\end{figure}

\begin{figure}[!h]%
	\includegraphics[scale=0.18]{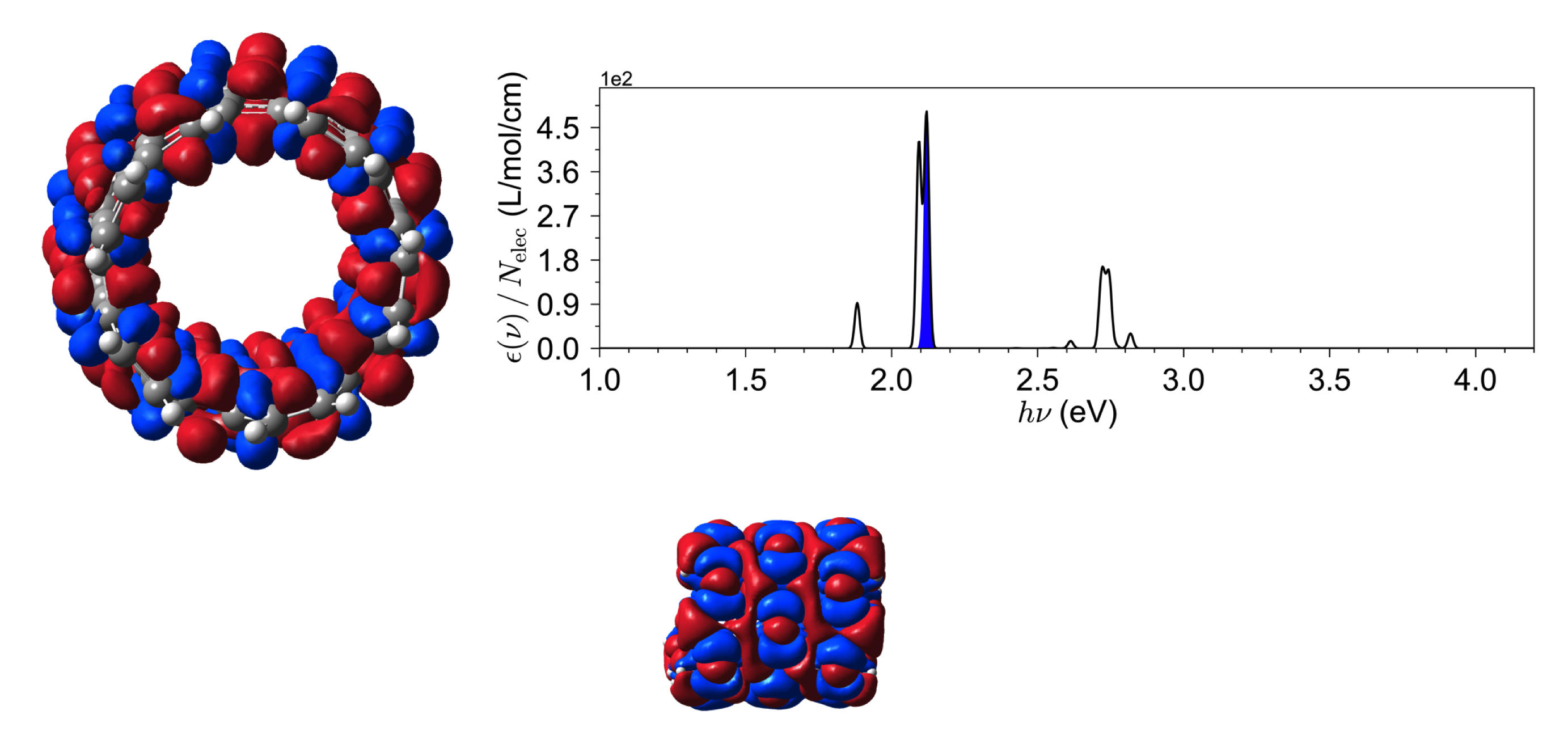}
	\caption{(Anomalous) NTO wave of the $E_{11}$ transition, highlighted in blue in the spectrum, for a finite, pristine (6,5) SWCNT with a length of $x=4$ cu ($\approx$ 1.0 nm).}
	\label{NTO_wave_04cu}
\end{figure}
\newpage

\phantom{This text will be invisible}
\begin{figure}[!h]%
	\includegraphics[scale=0.18]{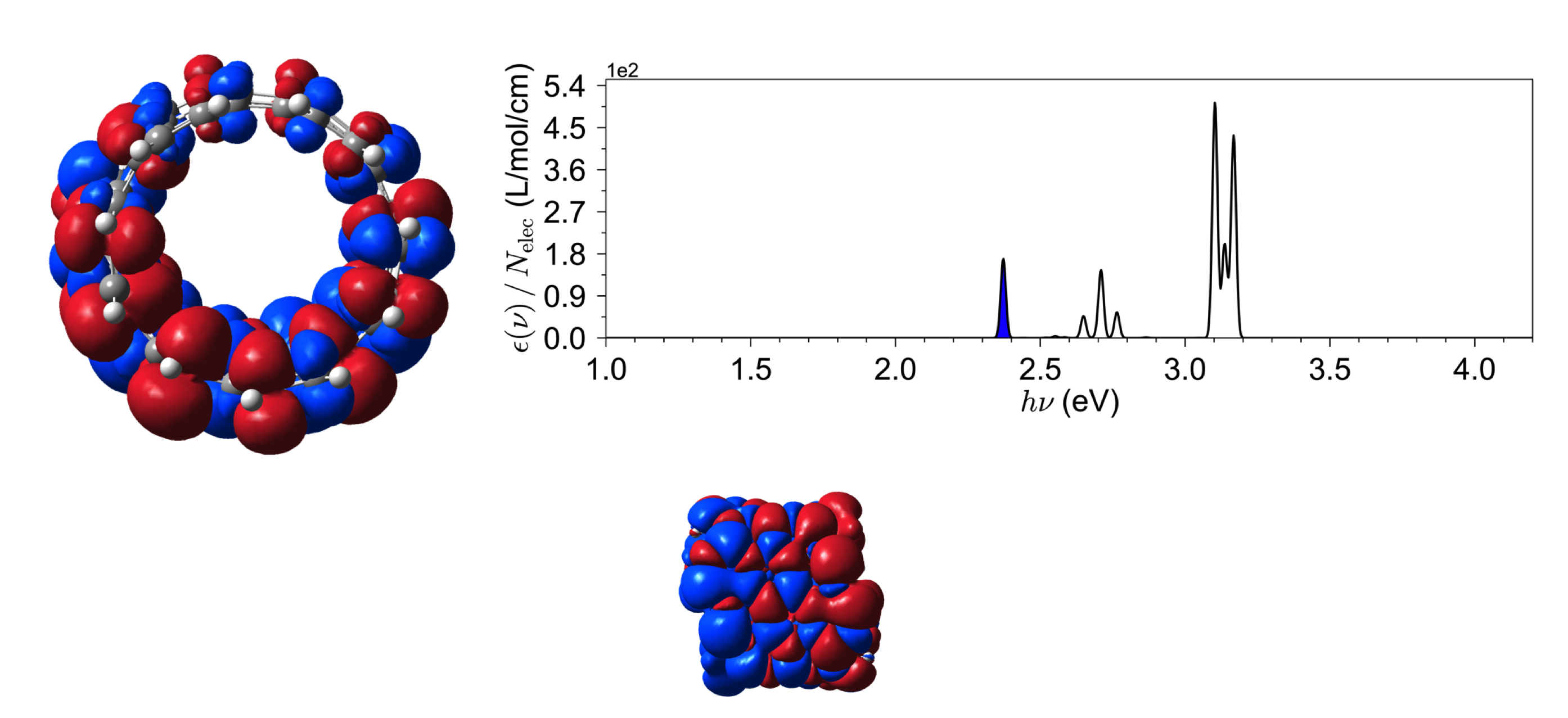}
	\caption{NTO wave of the $E_{11}$ transition, highlighted in blue in the spectrum, for a finite, pristine (6,5) SWCNT with a length of $x=3$ cu ($\approx$ 0.75 nm).}
	\label{NTO_wave_03cu}
\end{figure}

\begin{figure}[!h]%
	\includegraphics[scale=0.18]{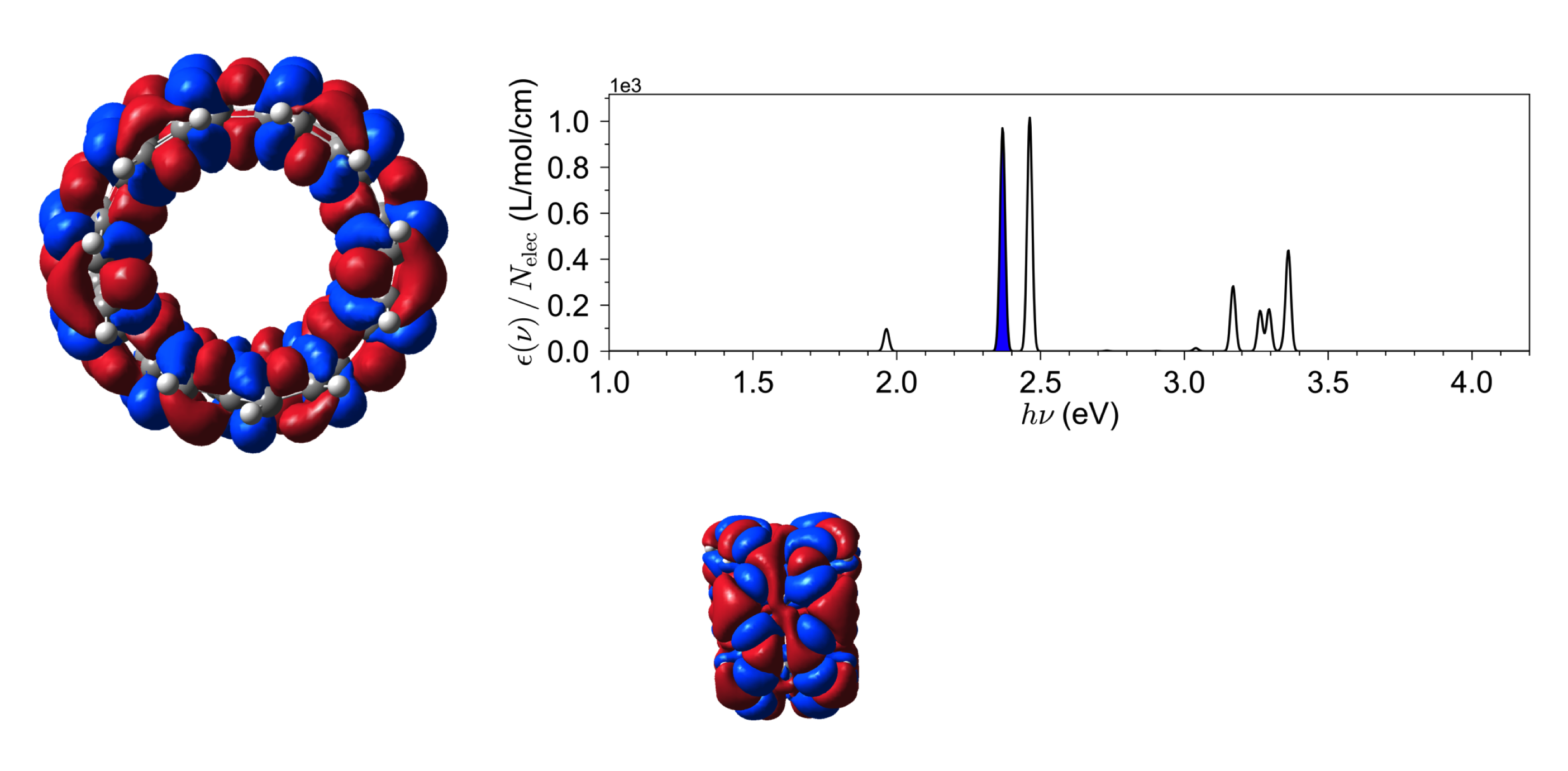}
	\caption{NTO wave of the $E_{11}$ transition, highlighted in blue in the spectrum, for a finite, pristine (6,5) SWCNT with a length of $x=2$ cu ($\approx$ 0.5 nm).}
	\label{NTO_wave_02cu}
\end{figure}
\newpage

\phantom{This text will be invisible}
\begin{figure}[!h]%
	\includegraphics[scale=0.18]{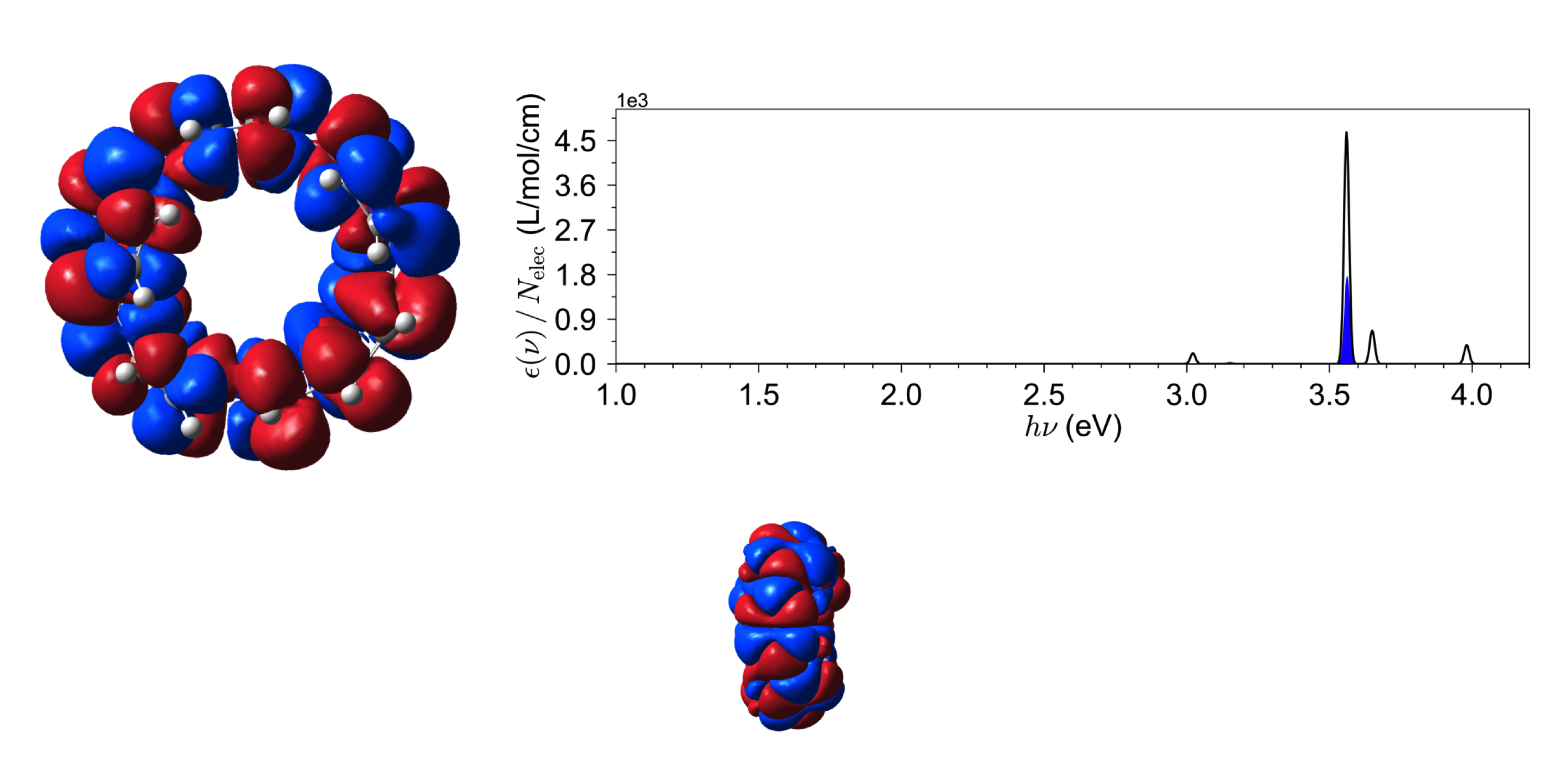}
	\caption{(Anomalous) NTO wave of the $E_{11}$ transition, highlighted in blue in the spectrum, for a finite, pristine (6,5) SWCNT with a length of $x=1$ cu ($\approx$ 0.25 nm).}
	\label{NTO_wave_01cu}
\end{figure}
\phantom{This text will be invisible}
\newpage

\section{Atomic coordinates of SWCNTs}
\begin{figure}[!h]
	\centering
	\includegraphics[scale=0.1]{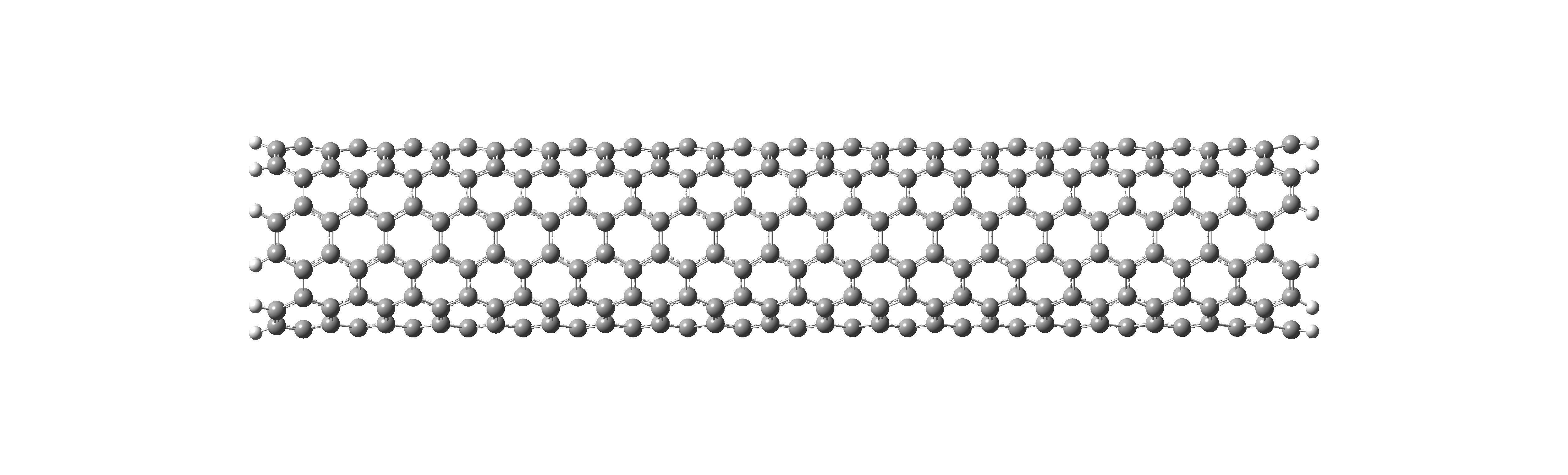}
	\caption{(6,6) SWCNT, $L\approx 4.5$ nm (36 cu), filename `6-6\_cnt\_36cu\_3-21G\_b3lyp\_optimized.xyz'}
	\label{6-6_36cu}
\end{figure}

\begin{figure}[!h]
	\centering
	\includegraphics[scale=0.1]{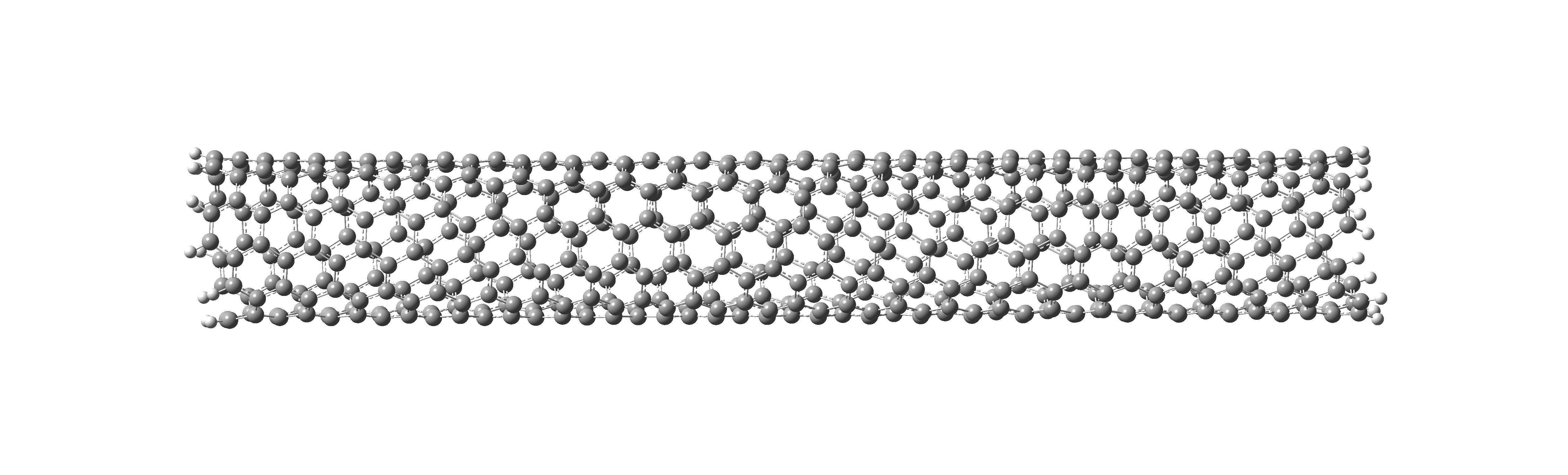}
	\caption{(6,5) SWCNT, $L\approx 5.5$ nm (22 cu), filename `6-5\_cnt\_22cu\_3-21G\_b3lyp\_optimized.xyz' }
	\label{6-5_22cu}
\end{figure}

\begin{figure}[!h]
	\centering
	\includegraphics[scale=0.1]{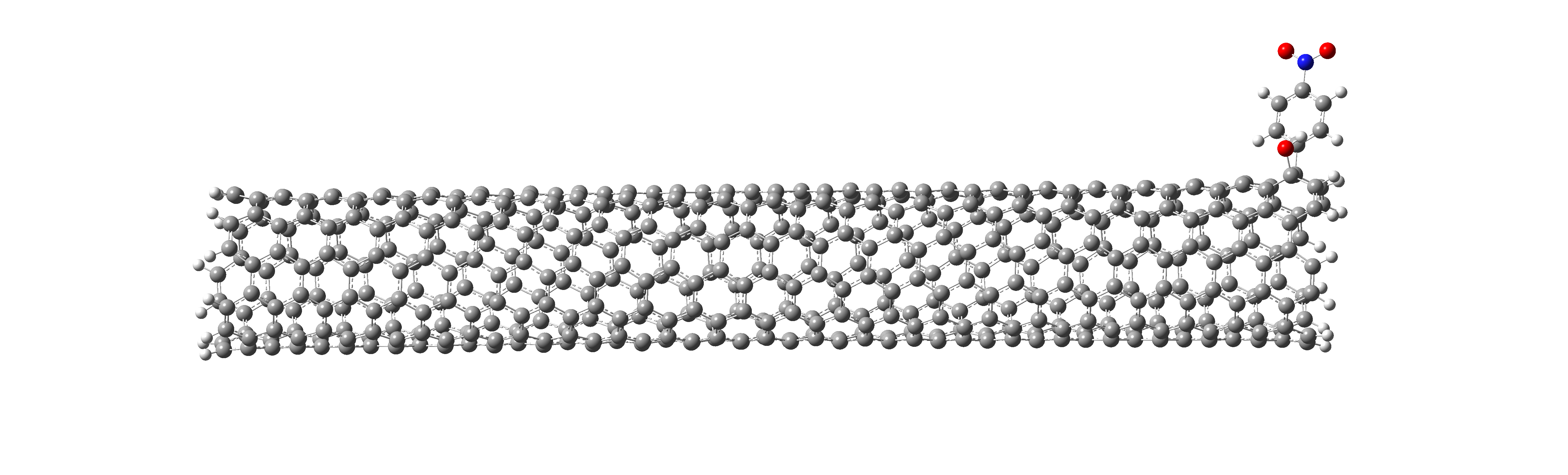}
	\caption{(6,5) FUN, $L\approx 5.5$ nm (22 cu), filename `6-5\_cnt\_aryl-NO2\_OH\_ortho90\_22cu\_3-21G\_b3lyp\_optimized.xyz' }
	\label{6-5_22cu_FUN}
\end{figure}

Molecular models of sample SWCNTs representative of those used in this study are provided as supplemental data files in .xyz format. These models are depicted in Figs.\,\ref{6-6_36cu}, \ref{6-5_22cu}, and \ref{6-5_22cu_FUN} and the corresponding filenames are included in the captions.

\phantom{This will be invisible}
\newpage

\bibliography{references.bib}